\documentclass{article}%
\usepackage{amsfonts}
\usepackage{array}
\usepackage{tabularx}
\usepackage{arydshln}
\usepackage{amsmath}

\DeclareMathOperator*{\argmin}{arg\,min}
\usepackage{amssymb}
\usepackage{authblk}
\usepackage{physics}
\usepackage{graphicx}%
\usepackage{caption}
\usepackage{hyperref}
\usepackage{subcaption}
\usepackage{multirow}

\usepackage{footmisc}
\providecommand{\U}[1]{\protect\rule{.1in}{.1in}}
\newtheorem{theorem}{Theorem}

\newtheorem{proposition}[theorem]{Proposition}
\newtheorem{remark}[theorem]{Remark}

\usepackage{xcolor}

\begin{document}

\title{A non-autonomous framework for climate change and extreme weather events increase in a stochastic energy balance model}
\author[1,2]{Gianmarco Del Sarto\thanks{gianmarco.delsarto@sns.it, gianmarco.delsarto@iusspavia.it}}
\author[1]{Franco Flandoli\thanks{franco.flandoli@sns.it}}

\affil[1]{Scuola Normale Superiore, Pisa, Italy}
\affil[2]{University School for Advanced Studies IUSS Pavia, Pavia, Italy}

\maketitle

\begin{abstract}
 We develop a three-timescale framework for modelling climate change and introduce a space-heterogeneous one-dimensional energy balance model. This model, addressing temperature fluctuations from rising carbon dioxide
  levels and the super-greenhouse effect in tropical regions, fits within the setting of stochastic reaction-diffusion equations. Our results show how both mean and variance of temperature increase, without the system going through a bifurcation point. This study aims to advance the conceptual understanding of the extreme weather events frequency increase due to climate change.

\end{abstract}

\section{Introduction}

The purpose of this paper is twofold: to describe a scheme for a three-timescale non-autonomous dynamical system suitable for the description of climate change and to present a new example of an energy balance model (EBM) with spatial structure. This model fits into our abstract framework and may contribute to a conceptual explanation of the increased frequency of extreme events. The non-autonomous nature of this example is crucial, making its connection to the foundational scheme particularly interesting. 

It is essential to emphasise two key aspects. First, the three-timescale scheme is designed to motivate the structure of the proposed EBM, including the adiabatic (in thermodynamics, a gas undergoes an adiabatic transformation during rapid expansion or compression; although fast, the transformation is slow enough that the gas stays in the state of statistical equilibrium and the transformation preserves the entropy of the gas. Here, and throughout the text, we use the term ‘adiabatic’ to refer to a variable that varies much more slowly than the timescale of the system of which it is a part) $\text{CO}_2$ variable, the presence of noise, and the system's timescale. Second, the EBM in question is a simplified climate model, representing the lowest level of complexity among climate models. As such, it can only provide qualitative insights into climate dynamics. Therefore, we deliberately avoid making quantitative comparisons between our model's results and real-world data.

\begin{figure}
\makebox[\linewidth][c]{%
\begin{subfigure}[b]{.42\textwidth}
\centering
\includegraphics[width=.99\textwidth]{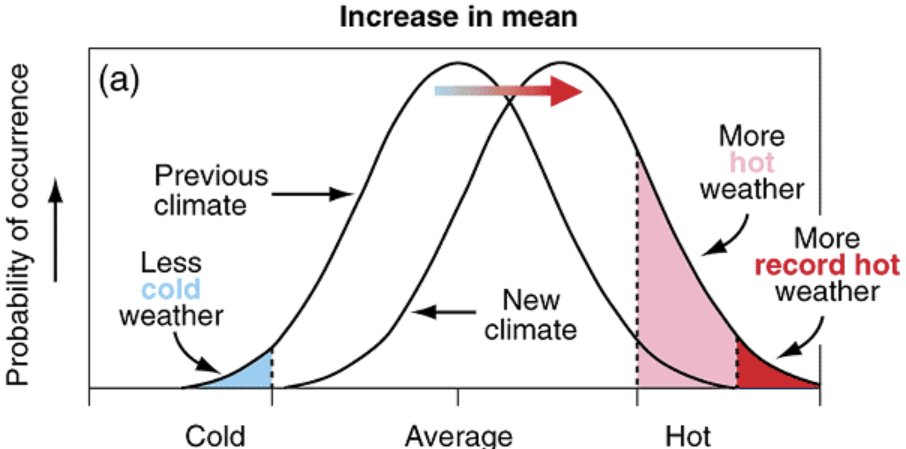}
%\caption{a test subfigure}
\end{subfigure}%
\begin{subfigure}[b]{.42\textwidth}
\centering
\includegraphics[width=.99\textwidth]{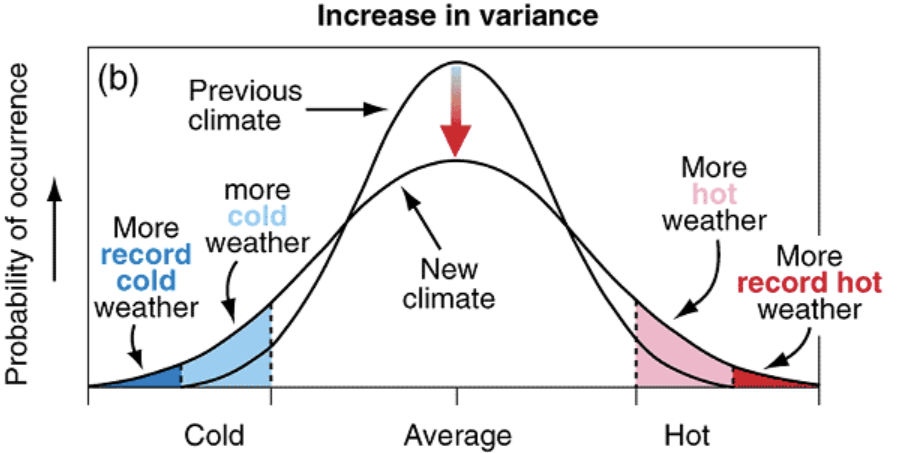}
%\caption{a test subfigure}
\end{subfigure}%
\begin{subfigure}[b]{.42\textwidth}
\centering
\includegraphics[width=.99\textwidth]{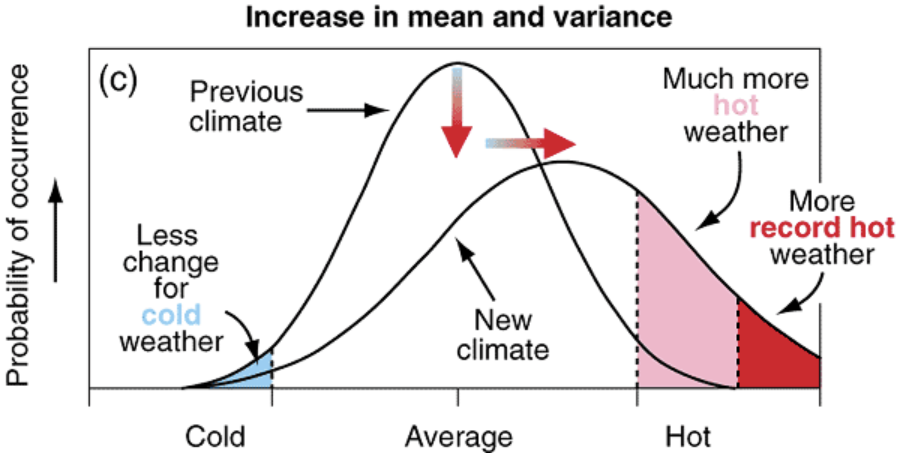}
%\caption{a test subfigure}
\end{subfigure}%
}
\caption{Schematic showing the effect on extreme temperatures when (a) the mean temperature increases, (b) the variance increases, and (c) when both the mean and variance increase for a normal distribution of temperature. Reproduced from the IPCC Report on Climate Change 2001: The Scientific Basis (\cite[Figure 2.32]{IPCC2001}).}
\label{fig: IPCC}
\end{figure}

\begin{figure}[!htb]
\centering
\begin{subfigure}{.49\textwidth}
  \centering
  \includegraphics[width=.95\linewidth]{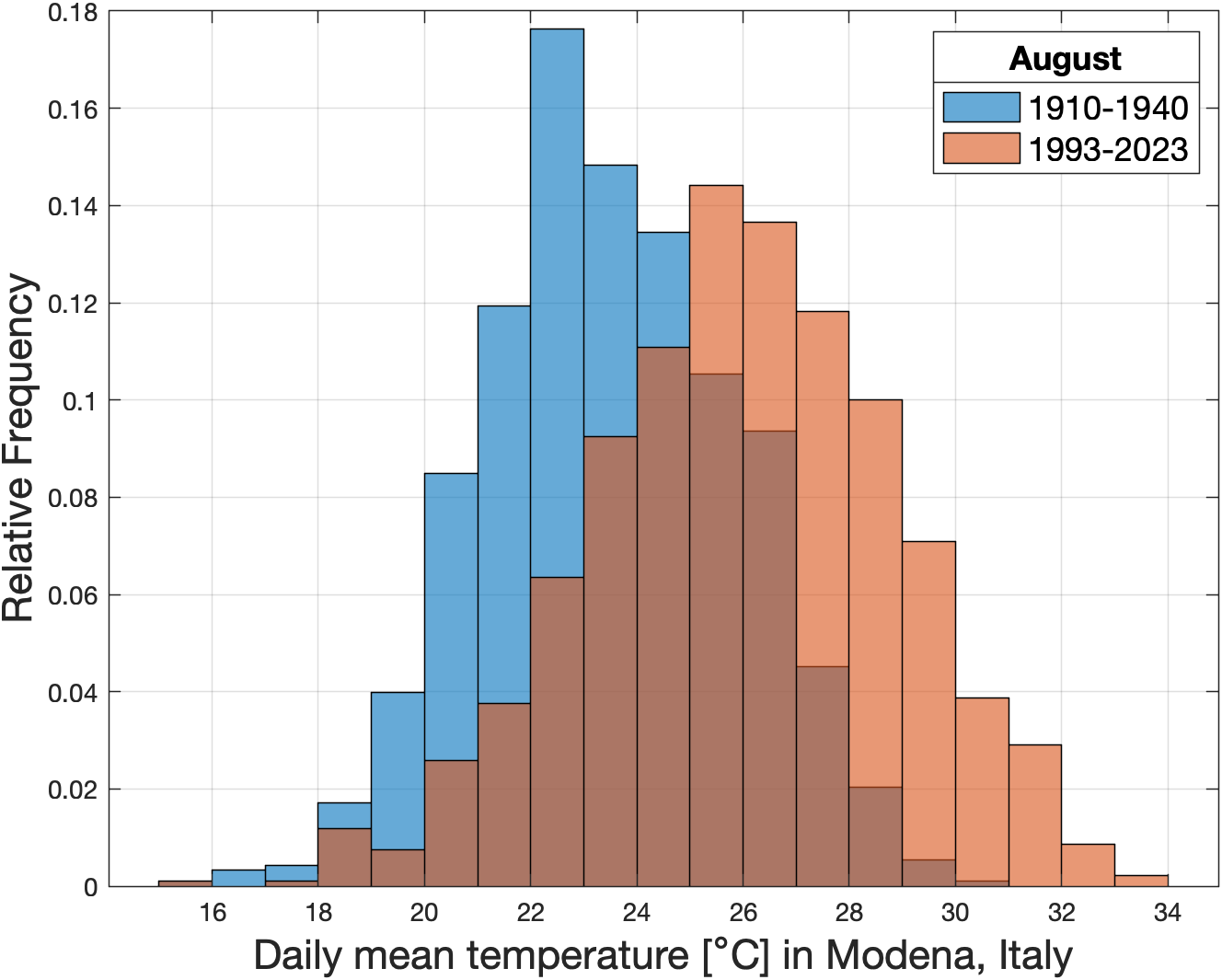}
  \caption{}
  \label{fig: data Modena}
\end{subfigure}
\begin{subfigure}{.49\textwidth}
  \centering
  \includegraphics[width = 0.65 \textwidth, height = 0.8\textwidth]{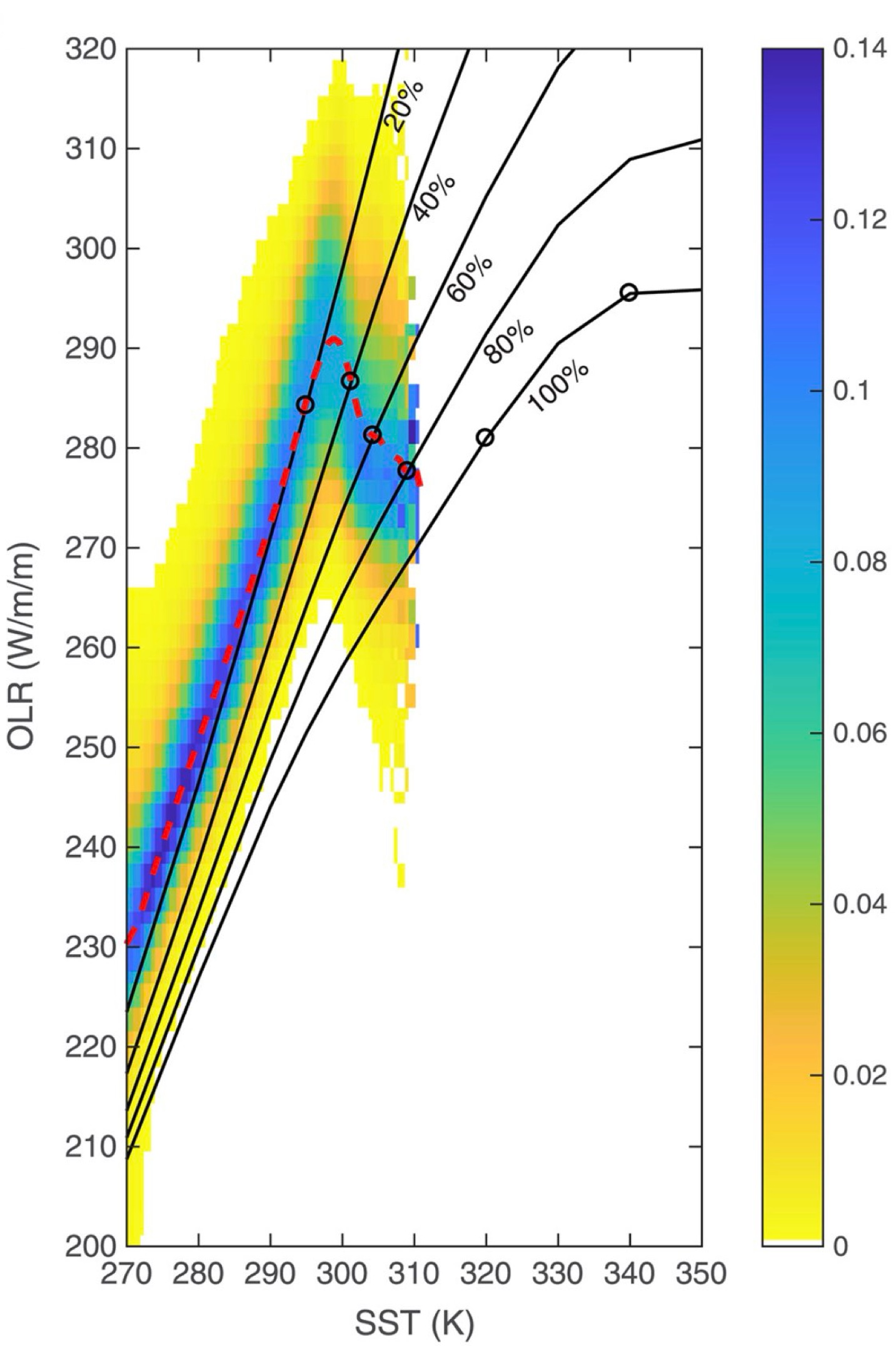}
  \caption{}
  \label{fig: OLR data}
\end{subfigure}%

\caption{(a) Daily temperature recorded in August in Modena, Italy. The blue histogram shows the values for the time period $1910-1940$ (blue), and the red histogram refers to $1993-2023$ (red). Data provided by the Osservatorio Geofisico di Modena \href{www.ossgeo.unimore.it}{www.ossgeo.unimore.it}. (b) Observational Outgoing Longwave Radiation (OLR) dependence on Sea Surface Temperature (SST), and $\text{SMART}$ (SMART is a dataset for OLR measurements used in \cite{Dewey2018}) OLR output for various humidity values. The red dashed line is the mean. Reproduced with permission from Geophysical Research Letters $45$, $19$ ($2018$). Copyright $2018$, John Wiley and Sons (\cite[Figure 2a]{Dewey2018}).}
\label{fig: data OLR tropics}
\end{figure}

From the viewpoint of climate change analysis, our goal is to provide a simplified model that demonstrates how temperature fluctuations increase with rising carbon dioxide ($\text{CO}_2$) concentrations. Probably the most dramatic example of climate change today is the increased frequency of extreme events. As pointed out by the IPCC Report (\cite{IPCC2001}), this can result from both an increase in the mean temperature and an increase in the variance of temperature, as shown in Figure \ref{fig: IPCC}, not only by the increase in the mean value but also by the increase in the variance of a climatic variable.
This phenomenon is not just speculative and can be observed by looking at real data, as can be done in Figure \ref{fig: data Modena}. It shows the histogram of the daily mean temperature recorded in August in two different periods, from $1910 $ to $1940$ in blue, and from $1993$ to $2023$ in red. In our opinion, being able to explain the joint increase of mean value and variance is a challenging and, at the moment, a completely open question. Indeed, a classical paradigm to explain the increase in the frequency of extreme events
is given by a dynamical system approaching a bifurcation point, a system
subject to random fluctuations, which are amplified near the bifurcation
point (\cite{Dakos2008,Scheffer2009,Kuehn2011,Ashwin2012, Lenton2012, Lucarini2020, Bernuzzi2023}). But the main question then is whether the Earth, nowadays, is close to a
bifurcation point or not, an issue which is speculated but not proven. Looking
to the far past, it is rather clear that bifurcations took place connecting a
moderate climate like our own today with glacial climates. But a clear
indication that now we are close to a new bifurcation point in the direction
of a warmer climate is missing. Except for certain data near the Tropics:
localised in space at that latitude, there is experimental evidence of a
potential bistability and therefore of the possibility of a fast transition to
a warmer climate (\cite{Dewey2018}). But far from the Tropics, the data are different. Therefore
we have developed a space-dependent model which incorporates this difference.
Seen globally, as an infinite dimensional dynamical system on the whole globe, the Tropics bistability does not add a new bifurcation to the model.

We focus on the class of EBMs, which give an elementary, yet useful, representation of Earth's climate by capturing the fundamental mechanisms governing its behaviour (\cite{ Budyko1969, Sellers1969, North1975, Ghil1976,Diaz1997,Lucarini2020, Cannarsa2023}). This type of models, which describes temperature evolution, assumes that it evolves according to the radiation balance of the budget, i.e. the difference in the radiation absorbed and emitted by the planet. Other key factors, such as insolation, atmospheric composition, and surface properties, are considered in EBMs in order to get a radiation budget as accurate as possible.

More in detail, we work with a Budyko-Sellers one-dimensional energy balance model (1D EBM), in which a diffusion term modelling meridional heat transport is added as a driver of temperature evolution (\cite{North2017}).  The novelty of our model consists in adding, for the first time, a particular phenomenon that may happen at the Tropics. Indeed, one of the key mechanisms to stabilise Earth's temperature is the Planck feedback. It consists of an increase of outgoing longwave radiation (OLR) as surface temperature increases, thanks to the Stefan-Boltzmann law. But there exist cases, and our model highlights one of them, where this feedback can fail, as in the super-greenhouse effect (SGE) (\cite{Emanuel2014, Beucler2016, Dewey2018}). This phenomenon is feedback between water vapour, surface temperature and greenhouse effect. Once the sea surface temperature or greenhouse gas concentration reaches a certain level, the increase in absorbed thermal radiation from the surface due to augmented evaporation with rising sea surface temperature (SST) outweighs the concurrent elevation in OLR, causing OLR to decline as SST increases.

From the mathematical viewpoint, the EBM proposed here is a non-autonomous stochastic partial differential Equation (SPDE) of reaction-diffusion type (\cite{DaPrato1996, DaPrato2014}). The non-autonomy stands in a slowly varying term corresponding to $\text{CO}_2$ concentration. We clarify the link between such a scheme and a more
classical adiabatic approach, where the slowly varying term is kept constant
and the long-term behaviour of the corresponding autonomous dynamical system is investigated. It is worth pointing out that the timescale of our model is intermediate between the daily timescale of weather and the centurial timescale of climate. It is sometimes called macroweather, and it is of the order of a month to one year. For this reason, the abstract framework of a two-timescale separation of a non-autonomous dynamical system is introduced to describe the macroweather-climate dichotomy, interpreting the weather as a faster timescale than macroweather which is averaged out by looking at the temperature on time interval of months.

This paper is organised as follows. In Section \ref{sec: the non-autonomous scheme with three timescales }, we introduce the non-autonomous framework for weather, macroweather, and climate, with a particular focus on the latter two. We outline the scale separation between them and provide their tentative definitions. This first, abstract part culminates in a link between macroweather and climate, justifying an adiabatic approach to studying the former in the presence of a non-autonomous dependence, such as $\text{CO}_2$ concentration in the atmosphere. In Section \ref{sec: Energy balance mdeols}, we begin by recalling the key concepts of EBMs, starting with the zero-dimensional version in Section \ref{sec: 3.1 }. We discuss why EBMs are suitable for describing macroweather and how they can explain the increase in global mean temperature (GMT) due to rising \( \text{CO}_2 \) concentrations. In Section \ref{sec: model formulation}, we introduce our new 1D EBM, detailing the parametrisation of all the terms of the parabolic partial differential equation (PDE). Section \ref{sec: deterministic properties of the model} describes the deterministic properties of the model, such as the existence of one or more steady-state solutions and their dependence on the \( \text{CO}_2 \) parameter. In Section \ref{sec: stochastic properties of the model}, we discuss the stochastic extension of the model and its properties, including the existence of an explicit invariant measure. Section \ref{sec: variance and extreme weather events increase } presents the main results of the second part. We demonstrate how our model predicts an increase in the variance of temperature fluctuations under \( \text{CO}_2 \) in the current climate configuration, without introducing new bifurcation points. We use both numerical tools and theoretical arguments to explain and understand this phenomenon. Finally, Section \ref{sec: conclusions} summarises our findings, while Appendix \ref{appendix: numerical scheme} and Appendix \ref{appendix: sturm liouville operators} detail the finite-difference scheme used for numerical simulations and some spectral properties needed to deduce the existence of the invariant measure for the stochastic problem.

\section{The non-autonomous scheme with three timescales}
\label{sec: the non-autonomous scheme with three timescales }

In Section \ref{sec: Energy balance mdeols} below, we introduce a stochastic EBM with suitable space
dependence, and a slowly varying parameter corresponding to $\text{CO}_{2}$
concentration, in order to investigate the time-change of fluctuations, as
also discussed in the Introduction. There are three different timescales
involved in this modelling. We could skip the first one by just mentioning \cite{Hasselmann1976}, and restrict ourselves to two timescales only, but at the same
time we want to insist on the non-autonomous structure of the modelling, hence
it is convenient to enlarge the discussion a bit.

In the modelling we have in mind, there are three timescales, called:

\begin{itemize}
\item weather, where variations are visible at the timescale of
hours/day (variables will be denoted by $V_{w}\left(  t\right)  $,
$T_{w}\left(  t\right)  $ etc.);

\item macroweather, where variations are visible at the timescale of
months/year (variables will be denoted by $V_{mw}\left(  t\right)  $,
$T_{mw}\left(  t\right)  $ etc.);

\item climate, changing at the timescale of dozens of years (probability
measures and their expected values characterise this level, still changing in time).
\end{itemize}
Subdivision and attribution of a precise timescale are not strict.

\begin{remark}
The reader will realise that, for the purpose of Section \ref{sec: Energy balance mdeols}, we could start
from Subsection \ref{section: macroweather and climate}. Let us then explain why we think that Subsections \ref{sec: the weather timescale}-\ref{sec: Hasselmann proposal}
are also very important. As already said, a main purpose of this work is
to identify possible explanations for an increase in variance, when a parameter
changes. In a sentence, the two main ingredients are a noise in the system
equations and a suitable nonlinearity which amplifies the variations of the
noise in a different way for different values of the parameter. The noise,
then, is crucial. Subsections \ref{sec: the weather timescale}-\ref{sec: Hasselmann proposal} are devoted to explaining its origin.
\end{remark}

\subsection{The weather timescale}
\label{sec: the weather timescale}

Following \cite{Hasselmann1976}, it is natural to model the weather scale by
deterministic equations, ordinary equations for simplicity of notation (but
the ideas are the same for PDEs); randomness can be introduced but it is not
strictly necessary, except maybe for a description of the uncertainty about
initial conditions and parameters, not included in the present discussion.
Following \cite{Hasselmann1976}, we distinguish the main physical variables in fast and
slow ones, according to a system of the form%

\begin{align*}
\partial_{t}V_{w} &  =f\left(  V_{w},T_{w}\right)  \\
\partial_{t}T_{w} &  =\epsilon g\left(  V_{w},T_{w},q\left(  \epsilon
t\right)  \right)  \\
&  q\left(  t\right)  \text{ slowly varying}%
\end{align*}
for a small $\epsilon>0$. Here $V_{w}$ changes in unitary time (corresponding
to hours/day), and $T_{w}$ varies very slowly (monthly, say). In addition, the
slow variable is influenced by a slow time-change of structure, described by
the time-varying parameters $q\left(  \epsilon t\right)  $. The function
$q\left(  t\right)  $ is assumed to be slowly varying, hence $q\left(
\epsilon t\right)  $ \textit{is super-slowly varying} (from here the three
timescales arise). 

The parameter $\tau=\frac{1}{\epsilon}$ corresponds to the typical reaction
time of the slow variables, measured in the unitary time of $V$. Appreciable
changes of $T$ happen in a time of order $\tau$, at the weather scale.

With great simplification, we may think that $V$ collects the fluid dynamic
variables (the fluid \textit{V}elocity plus other related variables), which
are very unstable and rapidly changing at the daily level, while $T$
represents \textit{T}emperature. In this case, the unit of time at the weather
level is of the order of hours, and $\tau=\frac{1}{\epsilon}$ is of the order of a
few months, hence e.g. of order 100. On the contrary, the time-change of
$\text{CO}_{2}$ concentration, call it $\tau_{\text{CO}_{2}}$, is of the order of a dozen of
years, hence e.g. of order 10000 in the weather scale. With these figures,
$q\left(  t\right)  $ has a relaxation time of $100$ and $q\left(  \epsilon
t\right)  $ of order $10000$.

\subsection{Macroweather timescale for $T$}
\label{sec: the macroweather timescale for T}

Then we change the scale and set%
\[
\widetilde{T}_{mw}\left(  t\right)  =T_{w}\left(  \frac{t}{\epsilon}\right)
\]
so that we observe variations of $\widetilde{T}_{mw}\left(  t\right)  $ in
unitary time. We call this the macroweather timescale. It holds%
\[
\partial_{t}\widetilde{T}_{mw}\left(  t\right)  =g\left(  V_{w}\left(
\frac{t}{\epsilon}\right)  ,\widetilde{T}_{mw}\left(  t\right)  ,q\left(
t\right)  \right)  .
\]
Let us recall that, at this timescale, $q$ varies very slowly. Let us look
for a simplification of this equation, where $V_{w}$ does not appear any more.

\subsection{The averaging approximation}
\label{sec: the averaging approximation}

Let us heuristically describe the averaging approximation, which can be made
rigorous under proper assumptions for suitable systems, see \cite{Freidlin2012}. 

At the integral level, we have%
\[
\widetilde{T}_{mw}\left(  t\right)  -\widetilde{T}_{mw}\left(  t_{0}\right)
=\int_{t_{0}}^{t}g\left(  V_{w}\left(  \frac{s}{\epsilon}\right)
,\widetilde{T}_{mw}\left(  s\right)  ,q\left(  s\right)  \right)  ds.
\]
If $t-t_{0}$ is small, let us use the reasonable approximation%
\begin{align*}
& \sim\int_{t_{0}}^{t}g\left(  V_{w}\left(  \frac{s}{\epsilon}\right)
,\widetilde{T}_{mw}\left(  t_{0}\right)  ,q\left(  t_{0}\right)  \right)
ds\\
& =\left(  t-t_{0}\right)  \frac{1}{t-t_{0}}\int_{t_{0}}^{t}g\left(
V_{w}\left(  \frac{s}{\epsilon}\right)  ,\widetilde{T}_{mw}\left(
t_{0}\right)  ,q\left(  t_{0}\right)  \right)  ds.
\end{align*}
Then, if $\epsilon<<t-t_{0}$, we heuristically invoke an ergodic theorem and
approximate%
\[
\sim\left(  t-t_{0}\right)  \int g\left(  v,\widetilde{T}_{mw}\left(
t_{0}\right)  ,q\left(  t_{0}\right)  \right)  \nu_{\widetilde{T}_{mw}\left(
t_{0}\right)  }\left(  dv\right)
\]
where $\nu_{\tau}\left(  dv\right)  $ is invariant for%
\[
\partial_{t}V=f\left(  V,\tau\right)  .
\]
Setting
\[
\overline{g}\left(  \tau,q\right)  =\int g\left(  v,\tau,q\right)  \nu_{\tau
}\left(  dv\right)
\]
we may write%
\[
=\left(  t-t_{0}\right)  \overline{g}\left(  \widetilde{T}_{mw}\left(
t_{0}\right)  ,q\left(  t_{0}\right)  \right)
\]
and then again approximate it to%
\[
\sim\int_{t_{0}}^{t}\overline{g}\left(  \widetilde{T}_{mw}\left(  s\right)
,q\left(  s\right)  \right)  ds.
\]
Hence we get the simplified model
\[
\partial_{t}\overline{T}_{mw}\left(  t\right)  =\overline{g}\left(
\overline{T}_{mw}\left(  t\right)  ,q\left(  t\right)  \right)  .
\]

\subsection{Hasselmann's proposal}
\label{sec: Hasselmann proposal}

However, in our case, this simplification is not realistic. If $\epsilon$ is of
order $\frac{1}{100}$, then $t-t_{0}$ is of order one, because we need the
validity of the approximation%
\[
\frac{1}{t-t_{0}}\int_{t_{0}}^{t}g\left(  V_{w}\left(  \frac{s}{\epsilon
}\right)  ,a,b\right)  ds\sim\int g\left(  v,a,b\right)  \nu_{a}\left(
dv\right)  .
\]
But on a time of order one, we observe variations of $\widetilde{T}_{mw}\left(
t\right)  $, we said above, hence the approximation%
\begin{equation*}
    \begin{split}
        & \phantom{a}\int_{t_{0}}^{t}g \left(  V_{w}\left(  \frac{s}{\epsilon}\right)
, \widetilde{T}_{mw}\left(  s\right)  , q\left(  s\right)  \right)  ds  
\sim \\
&\int_{t_{0}}^{t}g\left(  V_{w}\left(  \frac{s}{\epsilon}\right)
,\widetilde{T}_{mw}\left(  t_{0}\right)  ,q\left(  t_{0}\right)  \right)  ds
    \end{split}
\end{equation*}

is not so strict (on the contrary, it is excellent for the $q\left(  s\right)
\sim q\left(  t_{0}\right)  $ approximation).

\textit{We need to keep fluctuations into account at the macroweather scale}.
A phenomenological way (Hasselmann's proposal) is to replace the model above by%
\[
dT_{mw}\left(  t\right)  =\overline{g}\left(  T_{mw}\left(  t\right)
,q\left(  t\right)  \right)  dt+\sqrt{\epsilon}\sigma\left(  T_{mw}\left(
t\right)  ,q\left(  t\right)  \right)  \circ dW\left(  t\right)
\]
for a suitable "volatility" $\sigma$ (Stratonovich integral $\circ$ looks more
appropriate). In \cite{Hasselmann1976}, heuristic justifications are given, inspired
for instance the random displacements of a Brownian particle in a fluid of
molecules (which on their own are subject to a deterministic fast dynamics,
coupled with the slow deterministic dynamic of the bigger particle). The
Bremen school on Random Dynamical Systems and other research groups explored
for some time rigorous justifications for this proposal, but a final answer is
not known, see for instance \cite{Arnold2001, Kifer2001}. However, the observation of temperature time series at the
timescale of month-year clearly shows some form of stochasticity and thus
Hasselmann's proposal looks very appealing. 

For our purposes below, adhering to Hasselmann's proposal is essential, since
our results are the consequence of random perturbations of a nonlinear system
representing climate dynamics at the macroweather timescale, namely a
stochastic version of the EBM. Random perturbations are often accepted just
based on the generic justification of an unknown coupling with other segments
of the physical system (which at the end of the story is the reason also here,
namely the coupling with the fast variables) but Hasselmann's proposal is a more
precise explanation. 

Let us however advise the reader that we shall start, in our example below,
from a stochastic PDE for the temperature macroweather-scale, given a priori,
not derived precisely from the weather scale as described above. We want to
concentrate on the consequences of particular nonlinearities. Our model will
have the simplified form
\[
dT_{mw}\left(  t\right)  =\overline{g}\left(  T_{mw}\left(  t\right)
,q\left(  t\right)  \right)  dt+\sqrt{\epsilon}\sigma dW\left(  t\right)
\]
with constant $\sigma$.

\subsection{Macroweather and climate}
\label{section: macroweather and climate}

As announced at the end of the last subsection, our investigation starts from
an equation of the form (stochastic differential equation or SPDE)
\begin{align}
dT\left(  t\right)    & =\overline{g}\left(  T\left(  t\right)  ,q\left(
t\right)  \right)  dt+\sigma dW\left(  t\right)  \label{SDE di partenza}\\
& q\left(  t\right)  \text{ slowly varying}\nonumber
\end{align}
where we skip the subscripts but keep in mind that it is a macroweather
model, we have skipped the factor $\sqrt{\epsilon}$ but we shall choose a
small diffusion coefficient $\sigma$, and the nonlinear function $\overline
{g}$ will be chosen by means of typical arguments related to EBM's. The slowly
varying function $q\left(  t\right)  $ will describe the effect, in the model,
of slowly varying $\text{CO}_{2}$-concentration, appreciated on a timescale of
dozens of years. 

The climate is a collection of statistical information from the time series of
this model. If it were an autonomous system ($q\left(  t\right)  $ equal to a
constant), we would invoke invariant measures. Due to the time-change in
$\overline{g}$, we have to use the formalism of time-varying invariant
measures. However, at the simulation level, we approximate this time-varying
system by an adiabatic system parametrised by a parameter $q$:%
\[
dT\left(  t\right)  =\overline{g}\left(  T\left(  t\right)  ,q\right)
dt+\sigma dW\left(  t\right)
\]
and investigate its invariant measures, parametrised by $q$. The slow change
of statistics for the true non-autonomous system is mimicked by the change of
statistics when the parameter $q$ is changed.

Concerning precisely the concept of climate, let us introduce some formalism.
We again limit ourselves to stochastic differential equations (SDEs) on an Euclidean space $\mathbb{R}^{d}$ (e.g.
the space-discretisation of a stochastic partial differential equation, as in
our main example below) but the concepts can be widely generalised, see for instance \cite{Tonello2022} for a non-autonomous abstract random dynamical system framework related to the weather-climate dichotomy (that we improve
hereby introducing a third level, the macroweather). Call
$\Pr\left(  \mathbb{R}^{d}\right)  $ the set of probability measures on Borel
sets of $\mathbb{R}^{d}$. Consider the SDE (\ref{SDE di partenza}) on the full
real line of time. Assume that $W\left(  t\right)  $ is a $d$-dimensional
two-sided Brownian motion defined on a Probability space $\left(
\Omega,\mathcal{F},\mathbb{P}\right)  $ with expectation $\mathbb{E}$. Assume
that, for the Cauchy problem on the half line $\left[  t_{0},\infty\right)  $
with initial condition $T_{0}$ at time $t_{0}$, with arbitrary $t_{0}$ and
$T_{0}$, at least weak global existence and uniqueness in law holds, and
denote the solution by $T^{t_{0},T_{0}}\left(  t\right)  $, $t\in\left[
t_{0},\infty\right)  $; assume $T_{0}\mapsto T^{t_{0},T_{0}}\left(  t\right)
$ is Borel measurable from $\mathbb{R}^{d}$ to $\Pr\left(  \mathbb{R}%
^{d}\right)  $ endowed with the weak convergence of measures. 

For all $t_{0}<t$, we introduce the Markov semigroup $\mathcal{P}_{t_{0},t}$
mapping $\Pr\left(  \mathbb{R}^{d}\right)  $ into $\Pr\left(  \mathbb{R}%
^{d}\right)  $ defined by the identity%
\[
\int_{\mathbb{R}^{d}}\phi\left(  y\right)  \left(  \mathcal{P}_{t_{0},t}%
\nu\right)  \left(  dy\right)  =\mathbb{E}\int_{\mathbb{R}^{d}}\phi\left(
T^{t_{0},T_{0}}\left(  t\right)  \right)  \nu\left(  dT_{0}\right)
\]
for every $\nu\in\Pr\left(  \mathbb{R}^{d}\right)  $ and every bounded
continuous test function $\phi$ on $\mathbb{R}^{d}$. One can prove that
\begin{align*}
\mathcal{P}_{r,t}\mathcal{P}_{s,r} &  =\mathcal{P}_{s,t}\\
\mathcal{P}_{s,s} &  =Id.
\end{align*}
Moreover, one can link the Markov operator for equation (\ref{SDE di partenza}%
) to the Fokker-Planck equation%
\[
\partial_{t}f+\operatorname{div}\left(  \overline{g}\left(  \cdot,q\right)
f\right)  =\frac{\sigma^{2}}{2}\Delta f
\]
but we do not stress the rigorous results here.

The set of probability densities $\Pr\left(  \mathbb{R}%
^{d}\right)  $ is the state-space for the climate. In other words, any $\nu \in \Pr\left(  \mathbb{R}%
^{d}\right) $ is a (possible) state for the climate. Further, fixed the times $t_0 <t$, the operator $\mathcal{P}_{t_{0},t}$ defines the evolution, from $t_0$ to $t$, of the climate dynamics. Thus, we look for the climate concept inside the class of time-varying invariant measures which are invariant under the operator defining the climate dynamics, i.e. a family
\[
\left\{  \mu_{t}\right\}  _{t\in\mathbb{R}}\subset\Pr\left(  \mathbb{R}%
^{d}\right)
\]
such that
\[
\mathcal{P}_{s,t}\mu_{s}=\mu_{t}\qquad\text{for every }s\leq t\text{.}%
\]

\begin{remark}
    The concept of time-dependent invariant measure $\left \lbrace \mu_t \right \rbrace_{t \in \mathbb{R}}$ should not be confused with any solution of the Fokker-Planck equation. Similarly to the fact that, in many cases, the invariant measure $\mu$ of an autonomous system is the limit, as $t \to +\infty$, of the law of the solution $X_t^{x_0}$ starting at time $t = 0$ from the initial condition $x_0$, independent of $x_0$, the time-dependent invariant measure $\mu_t$ is expected to be, in many cases, the limit as $t_0 \to - \infty $ of the law of the solution $X_t^{t_0,x_0}$ starting at time $t_0$ from $x_0$, independently of $x_0$ (property that we could call "pull-back convergence to the equilibrium").

    Two simple illustrative examples are the Ornstein-Uhlenbeck equations with periodic or linear growth. For the periodic equation
    $$
    dX_t = - X_t dt + \sin(t)dt + dW_t
    $$
    the unique time-dependent invariant measure $\mu_t$ is the law, $2\pi$-periodic of the process
    $$
    X_t := \int_{-\infty}^t e^{-(t-s)} \sin(s) ds + \int_{- \infty}^t e^{-(t-s)}dW_s.
    $$
    For the linear growth equation (closer to our model with $\text{CO}_2$ increase)
    $$
    dX_t = - X_t dt + t dt + dW_t
    $$
    the unique time-dependent invariant measure $\mu_t$ is the law of the process
    $$
    X_t := \int_{-\infty}^t e^{-(t-s)}s ds + \int_{- \infty}^t e^{-(t-s)}dW_s.
    $$
    \end{remark}

Under suitable assumptions for the stochastic equations, there are results of
existence (easy, in particular relying on the existence of a compact global attractor) and also uniqueness (more difficult) for such invariant
families. This part of the theory is in progress. When the invariant measure
$\left\{  \mu_{t}\right\}  _{t\in\mathbb{R}}$ is unique, we call it "the climate". 

When uniqueness does not hold or it is not known, the idea could be to look
for families $\mu_{t}$ not only invariant but also with additional properties
of interest for physical sciences or other reasons. A typical one could be a
pull-back version of "convergence to equilibrium":%
\begin{equation}
\mu_{t}=\lim_{s\rightarrow-\infty}\mathcal{P}_{s,t}\lambda
\label{pull back Climate}%
\end{equation}
where $\lambda$ is a "natural" measure, as a rotation invariant centred
Gaussian measure on $\mathbb{R}^{d}$. For simplicity of understanding, the reader can assume that there is one and only one invariant family $\mu_t$ or one selected by
the pull-back property above.

If $q\left(  t\right)  $ varies very slowly, we expect that also $\mu_{t}$
varies accordingly, and thus an adiabatic approach to the numerical
computation of $\mu_{t}$ is reasonable, as already remarked above.

It is crucial to emphasise the following point. We believe that viewing climate as a time-dependent invariant measure, constructed in the pull-back sense, is not only a rigorous definition but also a physically meaningful one. Indeed, the current climate is the result of a long-term evolution that began in the distant past, where the dependence on the initial condition has been lost. This idea, together with the application to geophysical science of concepts from dynamical systems theory, developed in the 1990s (\cite{Crauel1994, Arnold1998}), began gaining traction approximately fifteen years ago (\cite{Ghil2008, Chekroun2011}).

\section{Energy balance models}
\label{sec: Energy balance mdeols}

\subsection{The macroweather timescale and the global mean temperature increase due to $\text{CO}_2$ concentration}

\label{sec: 3.1 }

EBMs are elementary climate models where, in the simplest form, the temperature of the planet evolves according to the balance of the radiation absorbed and emitted by the Earth (\cite{ Budyko1969, Sellers1969, North1975, Ghil1976,Lucarini2020}). Their ability to capture the essential dynamics of Earth's climate system while remaining computationally tractable makes EBMs valuable tools for understanding a wide range of climate phenomena, from the onset of ice ages to the impacts of greenhouse gas emissions on global temperatures (\cite{Dommenget2011,Bastiaansen2022}). There exists a spectrum of complexity for this kind of model, starting from the zero-dimensional (0D) case, moving to the one-dimensional (1D) case, and arriving at higher-dimensional models (\cite{North2017}). Before delving into the details of our 1D EBM with space heterogeneous radiation balance, we illustrate (i) why an EBM has a macroweather timescale, and (ii) why an increase of $\text{CO}_2$ concentration in a stochastic 0D EBM leads to an increase of GMT, but not the variance of the solution.

First, we consider a 0D EBM for the global mean temperature $T = T(t)$ which given a positive initial condition $T_0$, is an ODE of the form
\begin{equation}
    \begin{split}
        C \frac{dT}{dt} &= Q_0 \beta +q - A - B \cdot ( T - 273),\\
        T(0) & = T_0.
    \end{split}
    \label{eq: 0D EBM Budyko}
\end{equation}
In this model, the radiation emitted $R_e$ by the planet is assumed, according to Budyko empirical radiation formula (\cite{Budyko1969}), of the form
$$
R_e(T) = A+B \cdot (T-273 ) -q,
$$
where $A, B$ are positive constants that can be derived by a best-fit estimate with real data observations; $C>0$ denotes the heat capacity per square meter, while $Q_0$ and $\beta$ are respectively the global mean radiation and coalbedo. Lastly, the additive parameter $q>0$ the effect of $\text{CO}_2$ concentration on the radiation balance (\cite{North2017,Bastiaansen2022, DelSarto2024}). Denoting by $T_*$ the unique stable fixed point of the model, i.e.
$$
T_* = \frac{ Q_0 \beta -A +q}{B} + 273,
$$
the solution of Eq. \eqref{eq: 0D EBM Budyko} is given by
$$
T(t) = T_* + \left( T_0 - T_* \right) e^{-t /\tau_0},
$$
where $ \tau_0 = C/B$ is the relaxation time, i.e. the timescale at which a deviation from the equilibrium temperature is reabsorbed. Considering an all-land planet, it is reasonable to consider the heat capacity as half the heat capacity at constant pressure of the column of dry air over a square meter, as pointed out in \cite{North2017}, leading to
$$
 C = 5 \cdot 10^{7} \text{ J} \text{ K}^{-1} \text{ m}^{-2}.
$$
On the other hand, satellite data suggest (\cite{North2017,Graves1993})
$$
 B = 1.90 \text{ W } \text{m}^2 \text{ K}^{-1}.
$$
This leads to a relaxation time of the other of one month, as
$$
\tau_0 = \frac{C}{B} =  2.63 \cdot 10^7 \text{ s} \approx 30 \text{ days}.
$$
It is worth pointing out that the hypothesis of an all-land planet is a huge simplification of reality.
Indeed, the Earth's system has various components capable of storing heat efficiently, each with its own unique capacity (\cite{Lucarini2020}). The value for $C$ we have considered corresponds to the capacity of the atmospheric column. But, even considering a planet with a mixed-layer only ocean, the heat capacity would be $60$ times larger (see \cite{North2017}) i.e. in the order of a few years, remaining thus in the macro weather timescale.

Second, we force the model with a stochastic noise modelling the effect of fast terms with respect to the slow radiation balance terms of Eq. \eqref{eq: 0D EBM Budyko}. Denote by $(W_t)_{t \geq 0}$ a Brownian motion, and consider the SDE given by:
\begin{equation}
\begin{split}
    CdT_t &= \left(Q_0 \beta +q-A-BT_t \right) dt + \sigma dW_t, \\
    T(0) &= T_0,
    \end{split}
    \label{eq: stochastic EBM}
\end{equation}
where $\sigma >0 $ is the noise intensity. Denoting by $\Tilde{A}= Q_0\beta + q - A,$ the solution can be explicitly written, using a variation of parameters technique (\cite{Baldi2017}) as
$$
T_t = T_0 e^{-t/\tau_0} + \frac{\Tilde{A}}{\tau_0} \left( 1- e^{- t /\tau_0} \right) + \sigma \int_0^t e^{- (t-s)/\tau_0} dW_s.
$$
The solution is a Gaussian process with a mean value
\begin{equation}
\mathbb{E} \left[ T_t \right] = T(0) e^{- t / \tau_0} + \Tilde{A} \left( 1- e^{- t/\tau_0} \right),
\label{eq: mean OU stochastic EBM}
\end{equation}
and variance
\begin{equation}
    \begin{split}
        Var(T_t) &= Var \left(  \frac{\sigma}{C} \int_0^t e^{-(t-s)/\tau_0}ds\right) = \frac{\sigma^2}{C^2} \int_0^t e^{-2(t-s)/\tau_0}ds \\
        & = \frac{\sigma^2}{2BC} \left( 1-e^{-2 t / \tau_0} \right).
    \end{split}
    \label{eq: variance OU stochastic EBM}
\end{equation}
Further, the stochastic EBM in Eq. \eqref{eq: stochastic EBM} has a unique Gaussian invariant measure $\nu \sim \mathcal{N}(\mu_\nu, \sigma^2_\nu)$, whose mean $\mu_\nu$ and variance $\sigma^2_\nu$, can be obtained taking the limit for the time that tends to infinity in Eq. \eqref{eq: mean OU stochastic EBM} and Eq. \eqref{eq: variance OU stochastic EBM}, leading to
$$
\mu_\nu = \Tilde{A}= \frac{Q_0 \beta +q-A}{B},\qquad 
\sigma^2_\nu = \frac{\sigma^2}{2BC}.
$$
Thus, a change in the $\text{CO}_2$ concentration leads to a change in the mean value of the climate, i.e. the invariant measure, but not in its variability. Usually, the variability increase results from a critical transition, such as a saddle-node bifurcation, which arises in the model. This is the concept of bifurcation-induced tipping point, which leads to the critical slowing down behaviour of the system, resulting in an increase of variance and autocorrelation close to the bifurcation point (\cite{Dakos2008,Scheffer2009,Ashwin2012, Lenton2012}). However, the presence of a bifurcation point for the global scale climate is questionable, and the presence of bistable regimes for climate components, sometimes called tipping elements such for the Atlantic meridional overturning circulation or polar ice sheets, is localised in space. For all these reasons, in the next section, we will describe a new one-dimensional model with a local in-space change in the non-linear term that is able to explain the variance increase.

\subsection{Model formulation}
\label{sec: model formulation}
In this section, we propose a 1D EBM with space-dependent radiation balance with local bistability in the outgoing radiation term. The new term does not lead to the addition of a new bifurcation in the model, even if results in a non-linear change in the global mean temperature w.r.t. $\text{CO}_2$ concentration, in comparison with a linear increase in the case of the non-space dependent emitted radiation case. However, by adding a noise component to the model, we detect an increase of fluctuations over time for those values of $\text{CO}_2$ concentration in which the non-linear behaviour of GMT is triggered. We connect the increase in fluctuations over time, that we denote \emph{time variance}, to a local (in space) notion of \emph{relaxation time}. The latter indicator in a sense gives information about the local stability of a space point for, in our case, a global stable temperature configuration. 

The main characteristics of a 1D EBM are that the temperature $u= u(x,t)$, depending on time $t$ and space $x = sin(\phi)$, where $\phi \in [- \pi/2, \pi/2] $ denotes latitude, are that it evolves according to the diffusion of heat, and the planet energy balance (\cite{Budyko1969, Sellers1969, Ghil1976, Cannarsa2023}). Our model, which is an extension of the one considered in \cite{DelSarto2024}, assumes that the temperature $u$ satisfies the non-linear parabolic partial differential equation
\begin{equation}
    \begin{split}
        C_T\partial_t u &= \partial_{x} \left( \kappa (x) u_x \right) + R_a(x,u) - R_e(x,u;q),\;  x \in [-1,1], t \geq 0 \\
        u(x,0) &= u_0(x), \quad x \in[-1,1], \\
        u_x (-1,t) & = u_x(1,t) = 0, \quad t \geq 0.
    \end{split}
    \label{eq: 1D EBM}
\end{equation}
The heat capacity $C_T = 5 \cdot 10^{-7} \text{ J} \text{ K}^{-1} \text{ m}^{-2}$ is considered uniform over the whole planet, which we assume is an all-land planet, as the one presented at the beginning of Section \ref{sec: Energy balance mdeols}. The PDE is a reaction-diffusion type (\cite{Temam1997, Smoller2012}). Classical results can be used to prove the global existence and uniqueness of the solution, given a regular initial condition (\cite{Temam1997}). Further, it can be proved that $[0,+\infty)$ is an invariant region in the sense of \cite{Smoller2012}, as pointed out in \cite{DelSarto2024}. In the following, we are going to describe the terms governing its time evolution, while the values of the constants appearing in the parametrisations can be found in Table \ref{table: parameters Bastiaansen model}.

In scientific modelling, heat diffusion is often depicted assuming the planet as a thin shell. This leads to a non-constant diffusion coefficient of the form $\kappa (x) = D \cdot (1-x^2)$, with $D>0$, where the term $1-x^2$ arises due to the spherical setting (\cite{North2017}). However, this choice introduces difficulties for the mathematical treatment, resulting in degenerate problems (\cite{Floridia2014, Cannarsa2020}). For instance, proving the existence of steady-state solutions requires the use of weighted Sobolev spaces. To address this issue, we introduce a simplifying perturbation that removes the singularity at the border (\cite{DelSarto2024}). We consider the diffusion function given by:
$$
\kappa (x) = D \cdot (1-x^2) + \delta(x),
$$
with $D=0.3$ and $\delta \in C^\infty(-1,1)$, $\delta(x) = 0$ if $\abs{x} \leq \eta$, $\delta $ even and non-decreasing in $(0,1)$. The radiation absorbed by the planet, denoted as $R_a$, is the product of a spatially dependent solar radiation function $Q_0(x) = \Hat{Q}_0 \cdot (1-x^2)$, where $\Hat{Q}_0 >0$, and a temperature-dependent co-albedo $\beta = \beta(u)$. The co-albedo $\beta (u) = 1-\alpha(u)$, where $\alpha$ is parameterised by a smooth, non-increasing, bounded function (\cite{Bastiaansen2022, DelSarto2024})
\begin{equation}
\alpha(u) = \alpha_1 + \frac{\alpha_2 - \alpha_1}{2} \cdot \left[ 1 +\tanh{\left( K \cdot  \left( u- u_{ref}\right) \right)} \right],
    \label{eq: albedo}
\end{equation}
with $0 < \alpha_1 < \alpha_2 < 1$, $\alpha_1$ denotes ice albedo, $\alpha_2 $ denotes water albedo, $u_{ref} = 275 \text{ K}$, and $K$ is a parametrisation constant.

Third, we describe the modelling of the radiation emitted $R_e$, which is the main innovation of our model. Local bistability in the Outgoing Longwave Radiation (OLR) may arise in tropical regions due to a positive feedback loop involving surface temperature and moisture. This feedback, above a threshold of sea level pressure and GHG concentration, causes the atmosphere to become optically thick, thus reducing OLR (\cite{Dewey2018}).

\begin{figure}[!htb]
    \centering
    \includegraphics[width= 0.55 \textwidth]{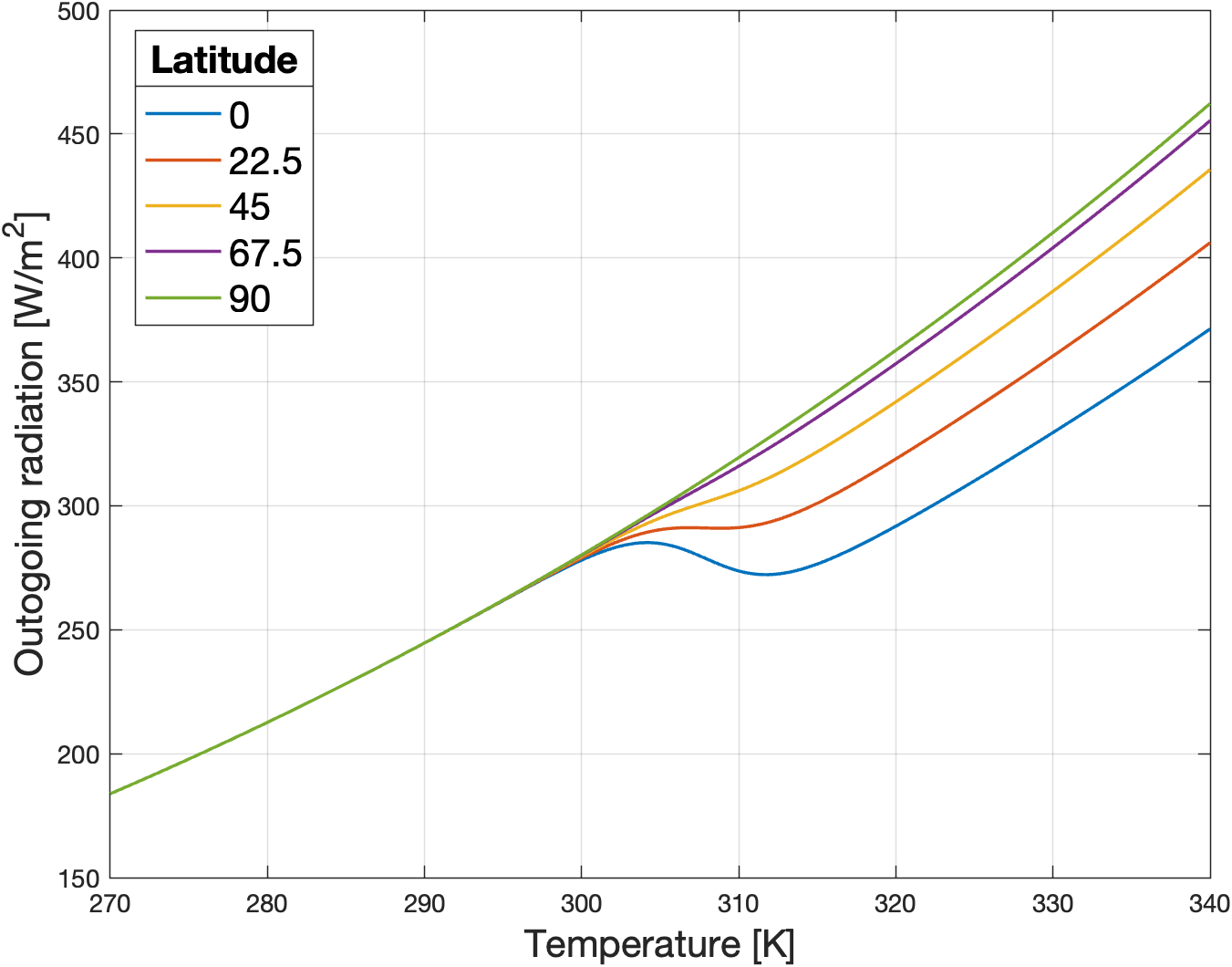}
    \caption{Representation of the Outgoing Longwave Radiation (OLR) $R_e = R_e(x,u;q) $ in \eqref{eq: OLR} for $q = 0$. Different colours are used to represent the dependence on the space. At the tropics, the OLR presents the bistability due to the super-greenhouse effect shown in Figure \ref{fig: data OLR tropics}. The bistable regimes progressively disappear as the space point moves to the poles, where the Stefan-Boltzmann law in \eqref{eq: OLR poles} is applied to parametrize OLR.}
    \label{fig: OLR model}
\end{figure}

To model this effect, we choose a space-dependent, latitude-symmetric Outgoing Longwave Radiation (OLR) of the form:
\begin{equation}
R_e(x,u;q) = q + \left| x \right| R_e^{pl}(u) + \left(1 - \left| x \right|\right) R_e^{eq}(u),
\label{eq: OLR}
\end{equation}
where \(R_e^{pl}\) and \(R_e^{eq}\) denote respectively the OLR at the pole and the equator. The convex combination between $R_e^{pl}$ and $R_e^{eq}$ Figure \ref{fig: OLR model} shows the space-dependent $OLR$ $R_e$, using different colours to represent distinct latitude points.
We define them as:
\begin{equation}
R_e^{pl}(u) = \varepsilon_0^{pl} \sigma_0 u\abs{u}^3,
\label{eq: OLR poles}
\end{equation}
where \(\varepsilon_0\) is the emissivity constant, \(\sigma_0\) is the Stefan-Boltzmann constant. On the other hand, at the poles, we reproduce the super-greenhouse effect by considering:
\begin{equation}
R_e^{eq}(u) = R_e^{pl}(u) g(u) + (1 - g(u)) \varepsilon_0^{eq} \sigma_0 u\abs{u}^3,
\label{eq: OLR equator}
\end{equation}
where \(g\) is a smooth transition function of the form
$$
g(u) = \frac{1}{1+ e^{\left(u-u_{ref}^{SGE}\right)K^{SGE}}},
$$
where $u_{ref}^{SGE} = 303.2$ K and $K^{SGE} = 0.36$ are respectively a reference temperature and a constant in the transition velocity involved in the SGE parametrisation. Further, \(\varepsilon_0^{eq} < \varepsilon_0^{pl}\) is an emissivity constant at the equator taking into account the OLR reduction due to the SGE. Note that the OLR at the poles and at the equator in Eq. \eqref{eq: OLR poles}-\eqref{eq: OLR equator} has been defined, just for mathematical convenience, in the physical meaningless range of negative Kelvin temperature.

Lastly, the additive positive parameter \(q\) models the effect of \(\text{CO}_2\) concentration on the energy budget (\cite{ Myhre2001,Bastiaansen2022}). A higher \(q\) leads to lower radiation emitted back to space from the surface. We emphasise the reason behind considering a constant value for the \(\text{CO}_2\) parameter \(q\). Greenhouse gas concentration can be regarded as constant at a macroweather, i.e., monthly, timescale, by avoiding seasonality insertion since it evolves on a much larger timescale. For instance, the \(\text{CO}_2\) concentration has increased by around \(47\%\) from \(1850\) to \(2020\), moving from \(284\) parts per million (ppm) to \(412\) ppm (\cite{IPCC2023}). In a first approximation, we assume that the spatial distribution of greenhouse gases is uniform over the globe. This assumption, although no longer the state of the art, was widespread at the turn of the century and is based on the well-mixing property of GHG. This means that since most GHGs, such as $\text{CO}_2$, have a large lifetime, many studies have been conducted using global average values for the spatial distribution (\cite{Myhre1998,Myhre2001, IPCC2001}).

\begin{table*}[!htb]
\caption{Parameters and constants appearing in the 1D EBM \eqref{eq: 1D EBM}.}
\begin{tabular}{ c c c }%{column = lcr}
\hline
 Symbol & Meaning & Value \\ [0.5ex] 
\hline
$D $ & Diffusivity constant & $0.45$ \\ 
 $\Hat{Q}_0$ & Mean solar radiation & $341.3 \; \text{W} \, \text{m}^{-2}$ \\
 $\varepsilon_0 ^{pl}$ & Emissivity at the poles & $0.61$ \\
 $\varepsilon_0 ^{eq}$ & Emissivity at the equator & $0.478$ \\
 $\sigma_0$ & Boltzmann's constant & $5.67 \cdot 10^{-8} \text{W} \,\text{m}^{-2} \, \text{K}^{-1}$ \\
 $\alpha_1$ & Ice albedo & $0.7$ \\
 $\alpha_2$ & Water albedo & $0.289$\\ 
 $K$ & Constant rate - albedo parametrisation & $0.1$ \\
 $K^{SGE}$ & Constant rate - SGE parametrisation & $0.1$ \\
 $u_{ref}$ & Reference temperature - albedo parametrisation & $273 \;\text{K}$ \\
  $u_{ref}^{SGE}$ & Reference temperature - SGE parametrisation & $303.2 \;\text{K}$ \\
  $C_T$ & Heat capacity & $5 \cdot 10^7 \;\text{J}\,\text{m}^{-2} \text{K}^{-1}$ \\
  [1ex] 
\hline
\end{tabular}
 \label{table: parameters Bastiaansen model}
\end{table*}

\subsection{Deterministic properties of the model}
\label{sec: deterministic properties of the model}

In this section, we describe the deterministic properties of our model, as the number of steady-state solutions and their stability. The new feature of our model is the rise of a strong non-linear increase of GMT with respect to the greenhouse parameter $q$, in the proximity of the model configuration describing the actual climate of the Earth.

In general, in dynamical system theory, huge information is given by the study of the steady-state solutions of the model, which are, in a sense, the long-time behaviour solutions. More specifically, in our model these solutions consist of the non-negative solutions $u = u(x)$ of the following elliptic PDE:
\begin{equation}
    \begin{split}
        0 &= \left( \kappa(x) u' \right)' + R_a(x,u) - R_e(x,u;q), \\
    0 &= u'(-1) = u'(1).
    \end{split}
    \label{eq: elliptic PDE}
\end{equation}
A common way to prove the existence of at least one stable steady-state solution involves a variational approach by studying the minimization problem
\begin{equation}
\inf \left\{ F_q (u) \mid u \in H^{1,2}(-1,1), \; u \geq 0 \right\}, 
\label{eq: variational problem}
\end{equation}
where:
\begin{equation}
F_q(u) = \int_{-1}^1 \mathcal{R}(x,u(x)) \, dx + \frac{1}{2} \int_{-1}^1 \left[ \kappa(x) u'(x) \right]^2 \, dx,
    \label{eq: F_q(u)}
\end{equation}
with \(\partial_u \mathcal{R} = R_e - R_a\) and \(H^{1,2}(-1,1)\) denoting the Sobolev space of order \(1\) and exponent \(2\) (\cite{ North1975, North1979, North1990, Brezis2011}). Applying the results from \cite[Theorem 1 and Theorem 3]{DelSarto2024}, it is possible to gain information about the existence and uniqueness of a steady-state solution.
\begin{proposition}
\begin{enumerate}
    \item[(i)] There exists an unique minimiser $u_* \in C^\infty([-1,1])$ of the variational problem \eqref{eq: variational problem}. Further, $u_*$ solves the elliptic problem \eqref{eq: elliptic PDE} and it is stable for the dynamics of the 1D EBM \eqref{eq: 1D EBM}.
    \item[(ii)] The map $q \mapsto \frac{1}{2} \int_{-1}^1 u_*(x) ds$ is non-decreasing.
\end{enumerate}
\label{prop: properties deterministic EBM}
\end{proposition}
Note how the second part of the previous result gives qualitative information on the behaviour of the GMT. Furthermore, if the diffusion coefficient is sufficiently large and the 0D EBM, obtained by removing the diffusion term and averaging in space the radiation balance, is bistable, it can be rigorously proven the existence of a second stable steady-state solution and a third unstable one (\cite{DelSarto2024}).

\begin{figure}[!htb]
\makebox[\linewidth][c]{%
\begin{subfigure}[b]{.42\textwidth}
\centering
\includegraphics[width=.99\textwidth]{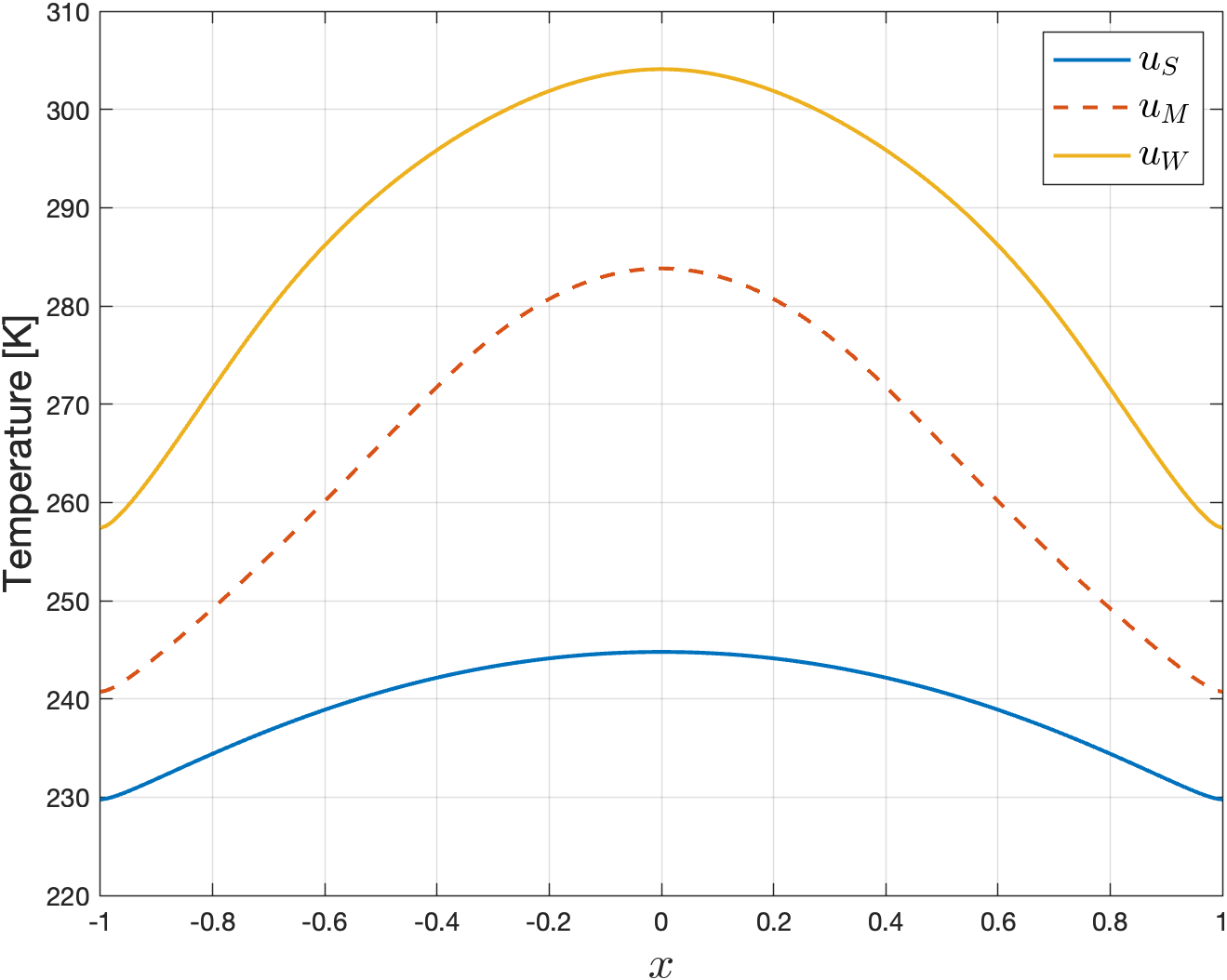}
\caption{}
\label{subfig: stationary solutions}
\end{subfigure}%
\begin{subfigure}[b]{.42\textwidth}
\centering
\includegraphics[width=.99\textwidth]{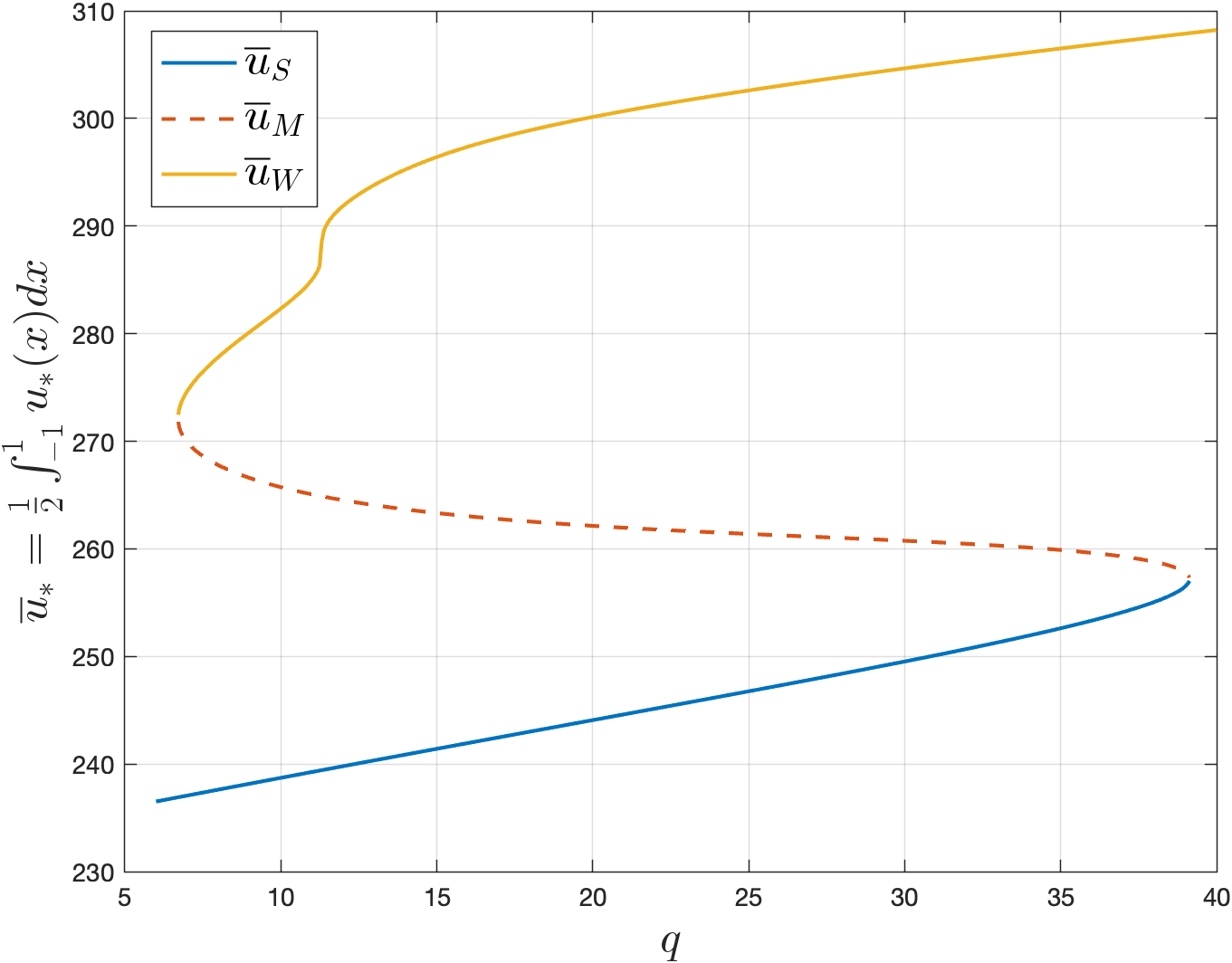}
\caption{}
 \label{subfig: bifurcation diagram}
\end{subfigure}%
\begin{subfigure}[b]{.42\textwidth}
\centering
\includegraphics[width=.99\textwidth]{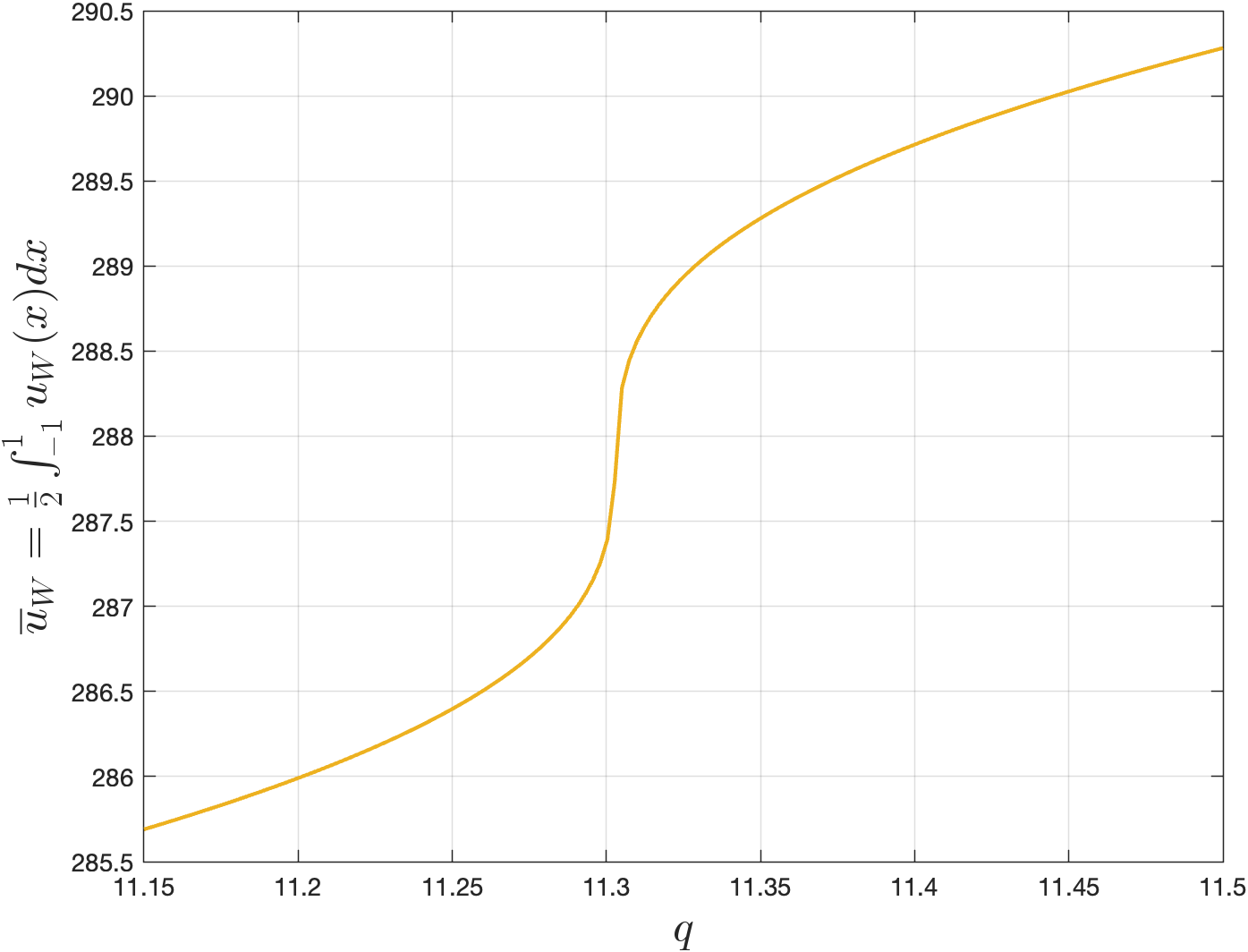}
\caption{}
\label{subfig: bif diagram u_w}
\end{subfigure}%
}
\caption{(a) Steady-state solutions for the 1D EBM \eqref{eq: 1D EBM} for $q = 11.3$. Solid lines denote stable solutions, dotted lines denote unstable solutions. The snowball solution $u_S$ is plotted in blue, the middle solution $u_M$ in red, and the warm solution $u_W$ in yellow. (b) Bifurcation diagram in the $(q,\Bar{u}_*)$ plan for the 1D EBM \eqref{eq: 1D EBM}, where $u_*$ denotes a steady-state solutions and $\Bar{u}_* = \frac{1}{2}\int_{-1}^1 u_*(x) dx$ is its global mean temperature (GMT). The $S$-shaped bifurcation diagram is characterized by the two classical saddle-node bifurcations around $q \approx 7$ and $q \approx 38$, and a non-linear (with respect to $q$) increase in GMT around $q \approx 22$. (c) Zoom of the bifurcation diagram around $q \approx 11.3$ and $u_W$. }
\label{fig: bifurcation diagrams}
\end{figure}

Given the non-linear nature of the problem, a rigorous demonstration of all properties of the model is not always possible. In this case, the problem can be overcome by using numerical simulations. In particular, for each fixed $q$, we numerically simulated the solutions of Eq. \eqref{eq: elliptic PDE}. As $q$ varies, we obtained, according to the large literature on this kind of model, that there can be either 1 or 3 stationary solutions. In the former case, the solution is stable. In the latter case, two solutions are stable, one describing a "snowball" configuration \(u_S\) for Earth's temperature with ice all over its surface. The other describes a warm climate \(u_W\), similar to the one in which we are living. Additionally, an unstable solution \(u_M\), whose average global mean temperature (GMT) lies between the GMT of \(u_S\) and \(u_S\) also arises. See Figure \ref{subfig: stationary solutions} for a graphical representation of the steady-state solutions in the bistable case.

In general, proving theoretically the results for such kinds of models is non-trivial due to the presence of nonlinear terms in the energy budget parametrisation.  The bifurcation diagram of the model in the \((q,\Bar{u}_*)\) plane, where \(u_*\) denotes a solution of the elliptic PDE, and \(\Bar{u}_* = \frac{1}{2} \int_{-1}^1 u_*(x) \, dx\) is its GMT, is depicted in Figure \ref{subfig: bifurcation diagram}. We highlight how the addition of a space-dependent OLR with tropics bistability does not introduce a new bifurcation. Thus, no new steady-state solutions are added or the stability of the already existing ones is altered. However, as highlighted in Figure \ref{subfig: bif diagram u_w}, a strong non-linear behaviour of the GMT for the warm solution $u_W$ appears for the value of $q$ in the neighbourhood of $q \approx 11.3$, a thing that does not happen if we remove the bistability in $R_e$ (\cite{DelSarto2024}). We aim to focus on that phenomenon, when a noise component, describing the weather effect, is added to the model. This is the main topic of the next sections.

\subsection{Stochastic properties of the model}
\label{sec: stochastic properties of the model}

As discussed in the previous paragraphs, the EBM presented in this work corresponds to a macroweather timescale. However, as far as it has been introduced, it lacks in taking into account the effect of fast components of Earth's system, such as atmospheric pressure, wind, and precipitation. In literature, this has been achieved elementary by adding a stochastic term, such as white noise, to the radiation budget (\cite{Hasselmann1976, Imkeller2001, Diaz2009, Diaz2022}). Note that this addition is necessary to describe the increase in the frequency of rare events. In fact, while a deterministic EBM is useful for obtaining information on global mean temperature, it is not able to investigate fluctuations around it due to all phenomena, such as weather, not included in the radiation balance of the model. 

We consider $ H = L^2(-1,1)$ and the stochastic EBM given by:
\begin{equation}
    \begin{split}
        du_t &=\left[ A u_t + R(x,u_t) \right]dt + \sigma dW_t,\\
        u_{\mid_{t = 0}} &= u_0
    \end{split}
    \label{eq: SPDE 1D EBM}
\end{equation}
where $(W_t)_t$ denotes a cylindrical Wiener process on $H$, $u_0 \in H$ is a non-negative initial condition and $A: D(A) \to L^2(-1,1)$ is the operator
\begin{equation}
\begin{split}
    D(A) &= \left \lbrace u \in H^2(-1,1) \; \mid \; u'(-1) = u'(1) = 0 \right \rbrace \\
    A u &=\left(\kappa(x) u' \right)'.
\end{split}
\end{equation}
We would like to apply the invariant measure theory for gradient systems to deduce the existence, and uniqueness of an invariant measure and its explicit formula. This can not be applied to the operator $A$ since it is invertible on $L^2(-1,1).$ For this reason, we consider the operator
\begin{equation}
\Tilde{A} = \lambda Id - A,
\label{eq: operatore tilde{A}}
\end{equation}
where $\lambda>0$ is a positive constant. By exploiting the Sturm-Liouville theory, it can be proved that $-\Tilde{A}$ is a self-adjoint, negative definite operator, with eigenvalue $0 < \lambda_1< \lambda_2< \cdots $, such that the trace of $ Tr\left[\left(-\Tilde{A}\right)^{\beta-1} \right] <+\infty $, for some $\beta \in (0,1)$. See Appendix \ref{appendix: sturm liouville operators} for a sketch of the proof of these facts. In this way, it is possible to deduce the following result (\cite{DaPrato2006, DaPrato2014, DelSarto2024}).
\begin{proposition}
Let $E = C([-1,1]).$ Then, there exists an unique $\mathbb{P}$-a.s. $E$-valued mild solution of the SPDE \eqref{eq: SPDE 1D EBM}. Further, there exists an unique Gibbs invariant measure $\Tilde{\nu}$, and $\Tilde{\nu} \ll \Tilde{\mu}$ with explicit formula
\begin{equation}
\Tilde{\nu}(du)= \frac{1}{Z} \exp \left(-\frac{2}{\sigma^2} \Tilde{I} (u)\right)  \Tilde{\mu}(du),
\label{eq: invariant Gibbs measure nu}
\end{equation}
where $\Tilde{\mu} \sim \mathcal{N} \left(0, -\frac{\sigma^2}{2} \Tilde{A}^{-1}\right)$ and
$$
\Tilde{I}_q(u) = \int_{-1}^{1} \mathcal{R}(x,u(x)) dx - \frac{\lambda}{2} \norm{u}_2^2.
$$
    \label{prop: existence, uniqueness mild solution, invariant measure}
\end{proposition}
We remark that the invariant measure $\Tilde{\nu}$ is the object, in our context of stochastic EBM, that in Section \ref{section: macroweather and climate} has been taken as the definition of climate. Then, it is worth pointing out the link between the invariant measure $\Tilde{\nu}$ and the functional $F_q$ building up the variational problem \eqref{eq: variational problem}. Indeed, at least formally, we can write the Gaussian measure $\Tilde{\mu}$ as
$$
\Tilde{\mu}(du) = \frac{1}{Z_1} \exp \left(- \frac{1}{2} \langle \mathcal{Q}^{-1} u,u \rangle \right) "du",
$$
where $Z_1$ is a normalization constant, $\mathcal{Q} = - \frac{\sigma^2}{2} \Tilde{A}^{-1}$ is the covariance operator of $\Tilde{\mu}$, $\langle \cdot, \cdot \rangle$ denotes the scalar product on $H =L^2 (-1,1)$ and $"du"$ is the formal notation for the Lebesgue measure on $H $. By an integration of parts, we deduce
$$
- \frac{1}{2} \langle \mathcal{Q}^{-1} u,u \rangle = - \frac{2}{\sigma^2} \left( \frac{\lambda}{2} \norm{u}_2^2 - \frac{1}{2} \langle \kappa(x) u', u' \rangle \right).
$$
Hence, substituting back the formal expression for $\Tilde{\mu}$ in Eq. \eqref{eq: invariant Gibbs measure nu}, we conclude
$$
\Tilde{\nu} (du) \propto \exp \left( - \frac{2}{\sigma^2}F_q(u)\right)  "du" .
$$
From this expression, we can deduce two important facts. First, the invariant measure of the stochastic EBM is concentrated on the global minimum points of the functional $F_q$. Second, the study of $\Tilde{\nu}$ and its spread depending on $q$ is as difficult as understanding how the functional $F_q$ changes. For this reason, in the next section, we will use numerical simulation to investigate how $\Tilde{\nu}$ changes, depending on the adiabatic parameter $q$, as the $\text{CO}_2$ lies in the critical interval in Figure \ref{subfig: bif diagram u_w}.

\subsection{Variance and extreme weather events increase}
\label{sec: variance and extreme weather events increase }

It is widely acknowledged that greenhouse gas emissions from human activities have led to more frequent and intense weather and climate extremes since the pre-industrial era, particularly temperature extremes (\cite{IPCC2023}). Despite numerous definitions proposed to assess what constitutes an extreme event, there is currently no universally accepted definition (\cite{Stephenson2008}). One commonly used definition considers an extreme weather event as one that exceeds a predefined threshold for a climate variable. 

As it is well understood that such occurrences become more likely as the variance of that climate variable increases, we consider the time variance as a proxy indicator of extreme weather events for our EBM setting. As explained in Section \ref{sec: stochastic properties of the model}, an explicit formula for the invariant measure of the stochastic EBM \eqref{eq: stochastic EBM} exists. However, due to the presence of a non-linear term inside the Gibbs factor in Eq. \eqref{eq: invariant Gibbs measure nu}, obtaining theoretical information is challenging. Thus, we rely on numerical simulations to capture the behaviour of the variance.

Specifically, given a fixed value of \( q > 0 \) and the warm steady-state solution \( u_W = u_W^{(q)} \), we numerically integrate the stochastic PDE:
\begin{equation}
    \begin{split}
        C_T\partial_t u &= \partial_{x} \left( \kappa (x) u_x \right) + R_a(x,u) - R_e(x,u;q)\\
        &+ \sigma dW_t,\quad  (x,t) \in [-1,1] \times [0,T], \\
        u(x,0) &= u_W(x), \quad x \in[-1,1], \\
        u_x (-1,t) & = u_x(1,t) = 0, \quad t \in [0,T].
    \end{split}
    \label{eq: 1D EBM time variance}
\end{equation}
The simulation runs for \( T = 500 \) years to capture the properties of the invariant measure around the warm climate \( u_W \), and we chose a noise intensity $\sigma = 0.2.$ The finite difference method applied to simulate the equation is detailed in Appendix \ref{appendix: numerical scheme}. Here, we describe what we mean by variance, its properties detected by numerical experiments, and our observations.

\begin{figure}[!htb]
\centering
\begin{subfigure}{.5\textwidth}
  \centering
  \includegraphics[width=.95\linewidth]{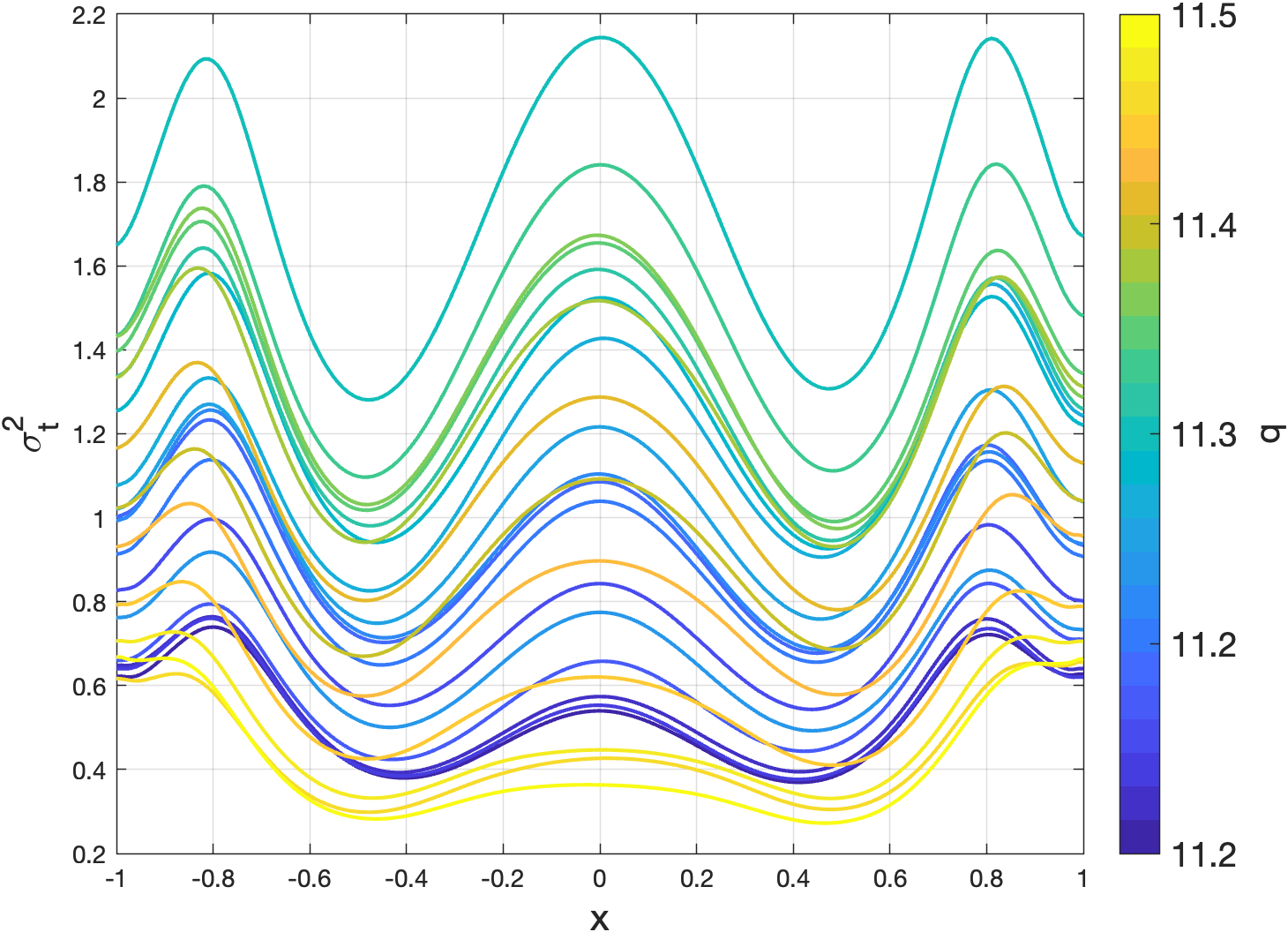}
  \caption{}
  \label{subfig: time variance}
\end{subfigure}%
\begin{subfigure}{.5\textwidth}
  \centering
  \includegraphics[width=.85\linewidth]{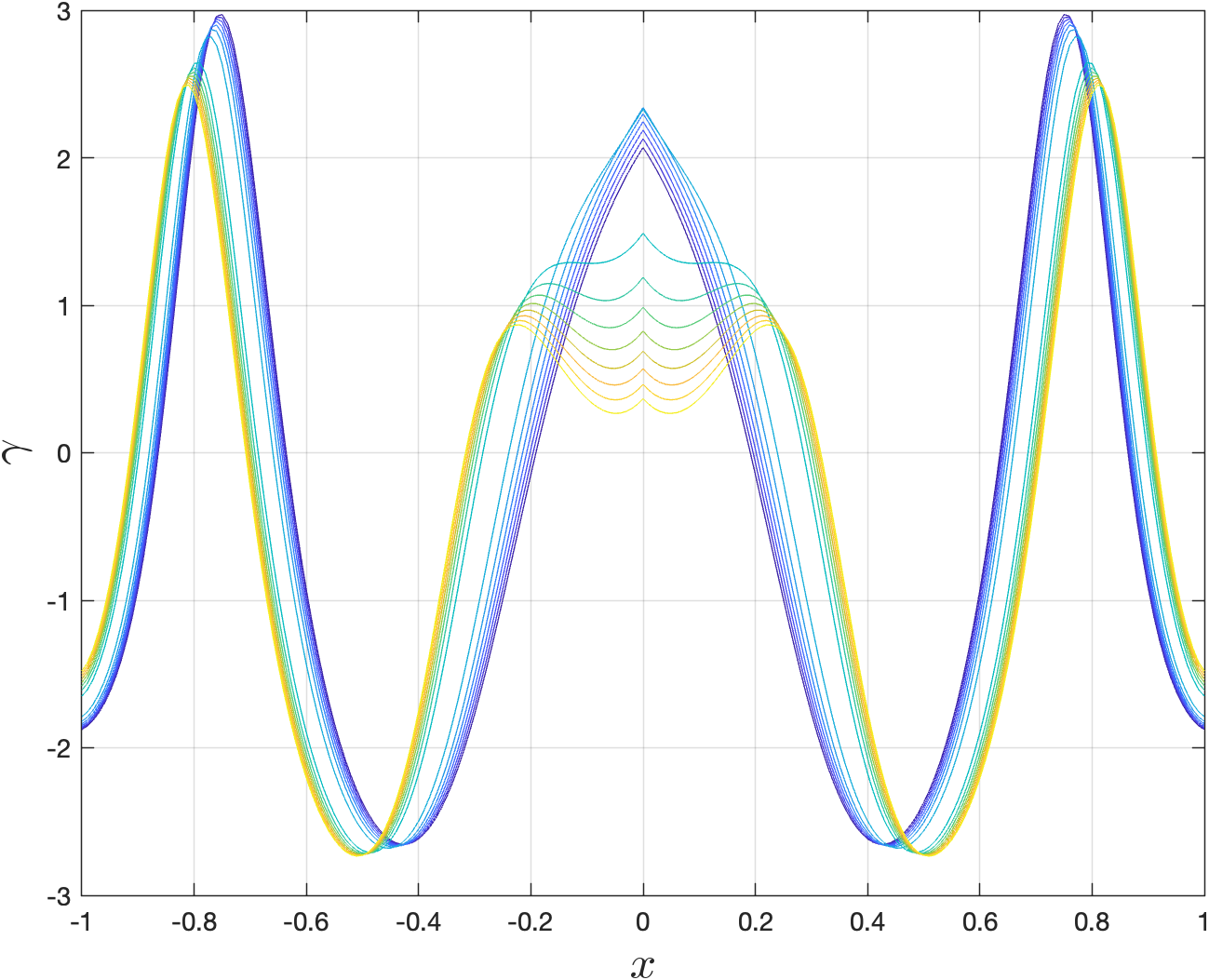}
  \caption{}
  \label{subfig: gamma}
\end{subfigure}
\caption{Indicators for the stochastic 1D EBM \eqref{eq: 1D EBM time variance} over the time interval \( [0,T] \), with \( T = 500 \text{ yrs} \). (a) Time-variance indicator \( \sigma^2_t \). Different colours represent different values of \( q \) for which the stochastic 1D EBM is simulated. (b) Local stability indicator \( \gamma \), using the same colour scheme as in subfigure (a).}
\label{fig: time variance and gamma}
\end{figure}

Given a space point \( x \in [-1,1] \) and a realization \( \omega \mapsto u(\omega) \) of the solution of the Eq. \eqref{eq: 1D EBM time variance}, we consider the variance of the process \( t \mapsto u_t(x,\omega) = u(x,t) \). We denote the numerical approximation of the solution by \( U = (u_{ij})_{ij} \), where \( u_{ij} = u(x_i, t_j) \), and \( (x_i)_{i=1,...,n} \) and \( (t_j)_{j=1,...,m} \) represent the spatial and temporal meshes, respectively, over the domain \( [-1,1] \) and \( [0, T] \). The time-variance is calculated as
\[
\sigma^2_t(x_i) = \frac{1}{m} \sum_{j = 1}^m \left(u_{ij} - \bar{u}_i\right)^2,
\]
where \( \bar{u}_i = \frac{1}{m} \sum_{j = 1}^m u_{ij} \). Figure \ref{subfig: time variance} illustrates the time-variance indicator, with different colours representing different values of the parameter \( q \). 

We observe two distinct behaviours of the variance. First, depending on \( q \), there is an increase in \( \sigma^2_t \), peaking around \( q \approx 11.3 \), followed by a decrease. This peak corresponds to the \( q \) value that results in the largest increase in GMT, as shown in Figure \ref{subfig: bif diagram u_w}. Second, as expected, for a given value of \( q \), the spatial profile of \( \sigma^2_t \) is symmetric with respect to \( x = 0 \). Additionally, three local maximum points emerge one at \( x = 0 \) and the others at \( |x| \approx 0.8 \). Our interpretation is that \( \sigma^2_t \) identifies the highest fluctuating regions as those where the freezing water temperature is crossed (sub-arctic regions, \( |x| = 0.8 \)), or where the SGE triggers bistability (tropical areas, \( x = 0 \)).

To support this interpretation, it is useful to introduce the local stability indicator
\[
\gamma(x) = \partial_u R(x,u_W(x)),
\]
where $R(x,u) = R_a(x,u) - R_e(x,u;q)$. If the model were a simple Ordinary Differential Equation (ODE) without the coupling diffusion term, \( \gamma(x) \) would be, up to a positive constant depending on $C_T$, the eigenvalue obtained by linearizing the reaction term around the stable steady-state temperature \( u_W(x) \). Positive values of \( \gamma(x) \) indicate local instability, while negative values indicate stability if the diffusion term is removed. Figure \ref{subfig: gamma} presents the local stability indicator \( \gamma \), with colours denoting different values of the parameter \( q \). Notably, there is a clear relationship between time-variance and the local stability indicator: regions with high time-variance \( \sigma^2_t \) coincide with areas of positive \( \gamma \). This explains the spatial profile of \( \sigma^2_t \) for a fixed value of \( q \).

The reason for \( \gamma(x) > 0 \) at some points is due to the presence of the diffusion term. If \( \kappa \equiv 0 \), then \( u_W \) is not a minimiser of \( F_q \), but instead solves the problem
\begin{equation}
u_W(x) = \argmin_{u \geq 0} \mathcal{R}(x,u),
\label{eq: argmin u_* case kappa =0}
\end{equation}
where $\partial_u\mathcal{R} = R_e- R_a$. Here, \( \partial_u \mathcal{R}_{\mid (x, u_W(x))} = 0 \), and since \( u_W(x) \) is a minimum point, it follows \( \partial^2_u \mathcal{R}_{\mid (x, u_W(x))} = \partial_u R _{\mid (x, u_W(x))}  \leq  0 \). But with \( \kappa(x) > 0 \), \( u_W(x) \) no longer satisfies Eq. \eqref{eq: argmin u_* case kappa =0}, but instead is a minimiser of \( F_q \), where the diffusion term appears in the second term on the right-hand side of Eq. \eqref{eq: F_q(u)}. This term, which intuitively minimises the temperature gradient, allows for some points to fluctuate around a mean value that would be unstable if \( \kappa \equiv 0 \).

To understand why, given \( x \in [-1,1] \), the map \( q \mapsto \sigma^2_t(x) = \sigma^{2,(q)}_t(x) \) increases until around \( q \approx 11.3 \) and then decreases, we offer the following explanation. As \( q \) increases, the temperature \( u_W(x) \) of the warm climate increases, as can be checked by numerical simulations and partially understandable from Proposition \ref{prop: properties deterministic EBM}. This leads the temperature, especially in the tropical area, to approach the region where instability due to the SGE arises. Once the tropical temperature surpasses this region, the instability, and thus the variance, decreases. We support our statements as follows.

\begin{figure}
\makebox[\linewidth][c]{%
\begin{subfigure}[b]{.42\textwidth}
\centering
\includegraphics[width=.99\textwidth]{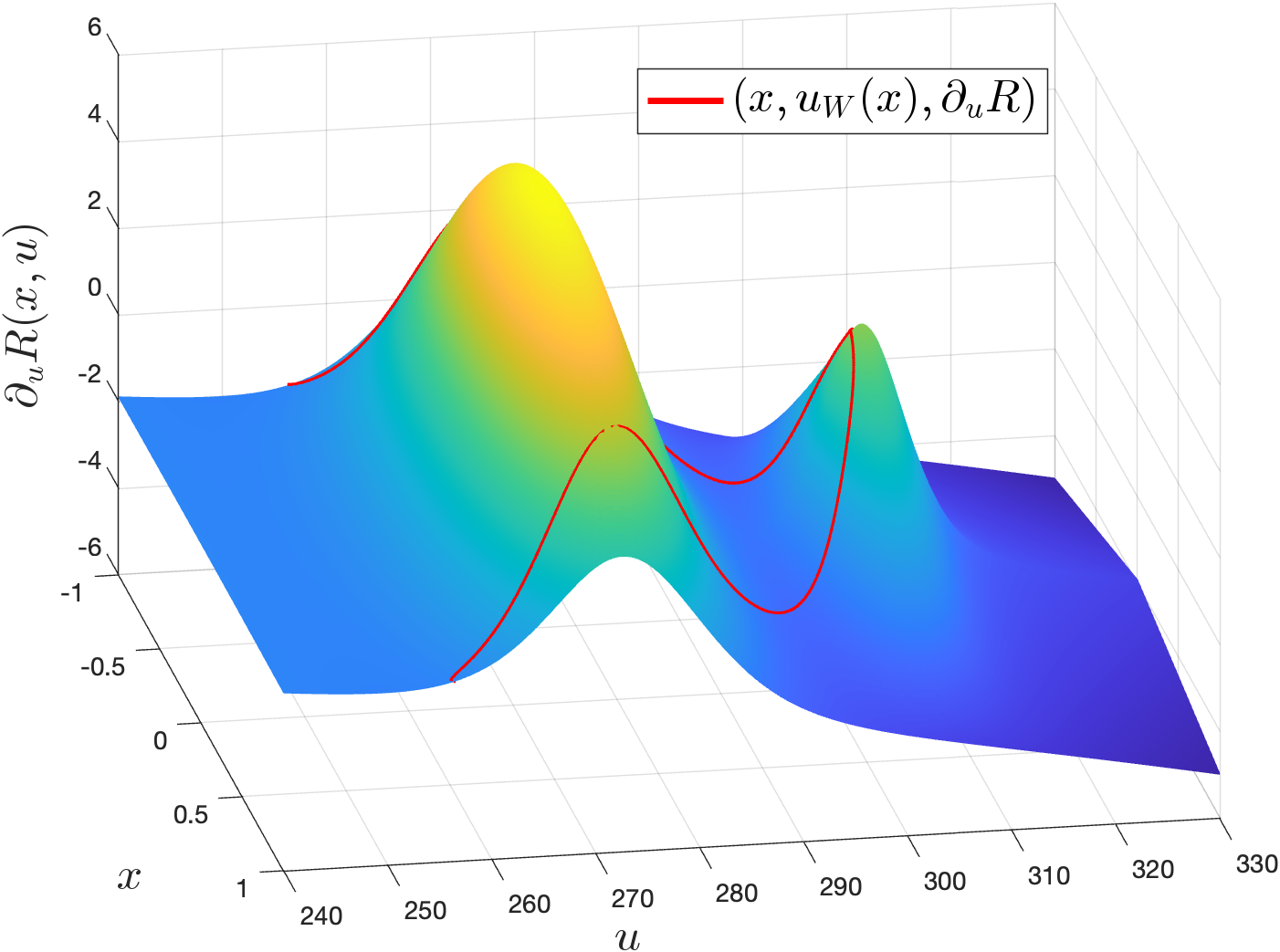}
\caption{$q = 11.21$}
\end{subfigure}%
\begin{subfigure}[b]{.42\textwidth}
\centering
\includegraphics[width=.99\textwidth]{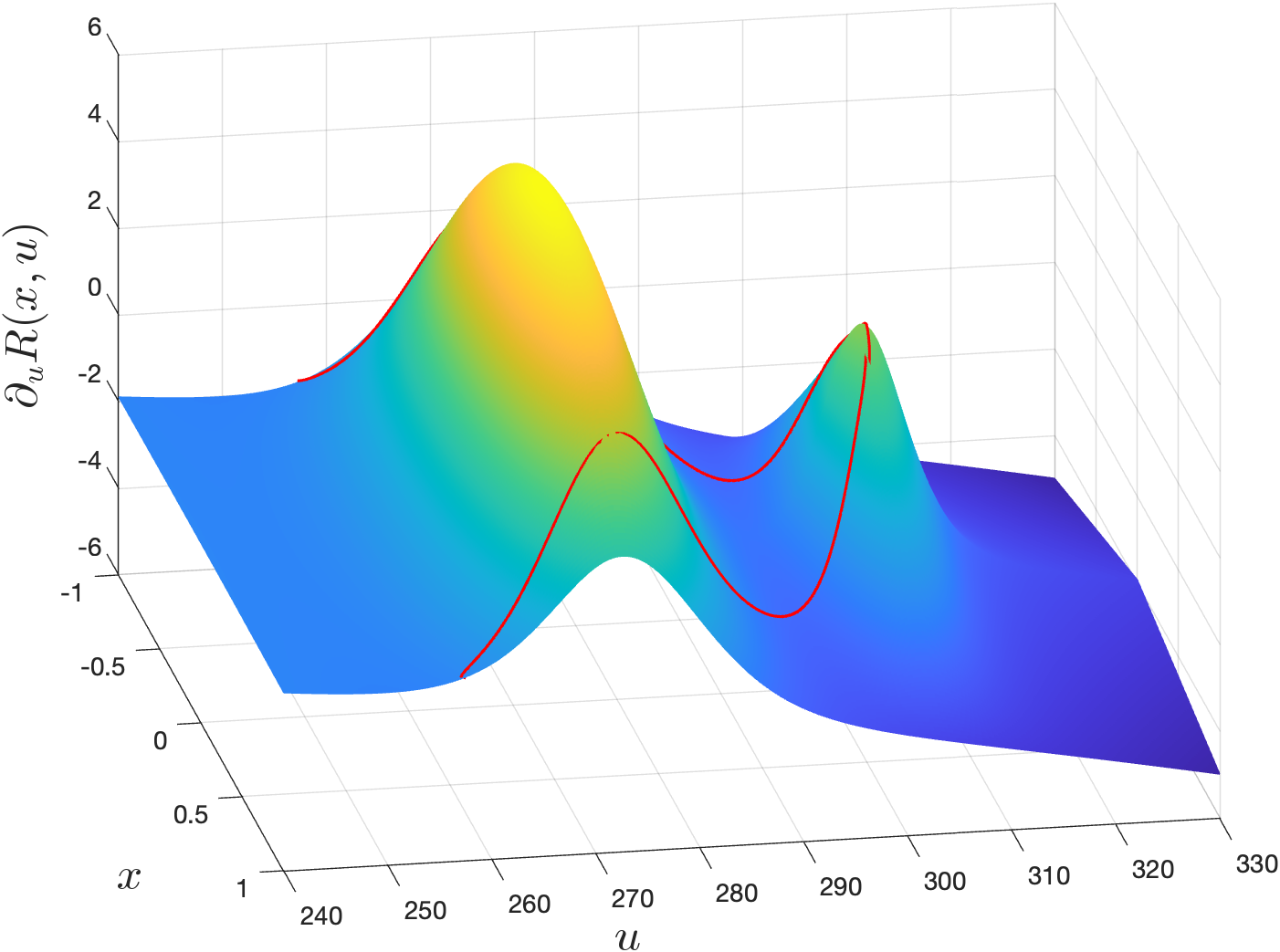}
\caption{$q = 11.29$}
\end{subfigure}%
\begin{subfigure}[b]{.42\textwidth}
\centering
\includegraphics[width=.99\textwidth]{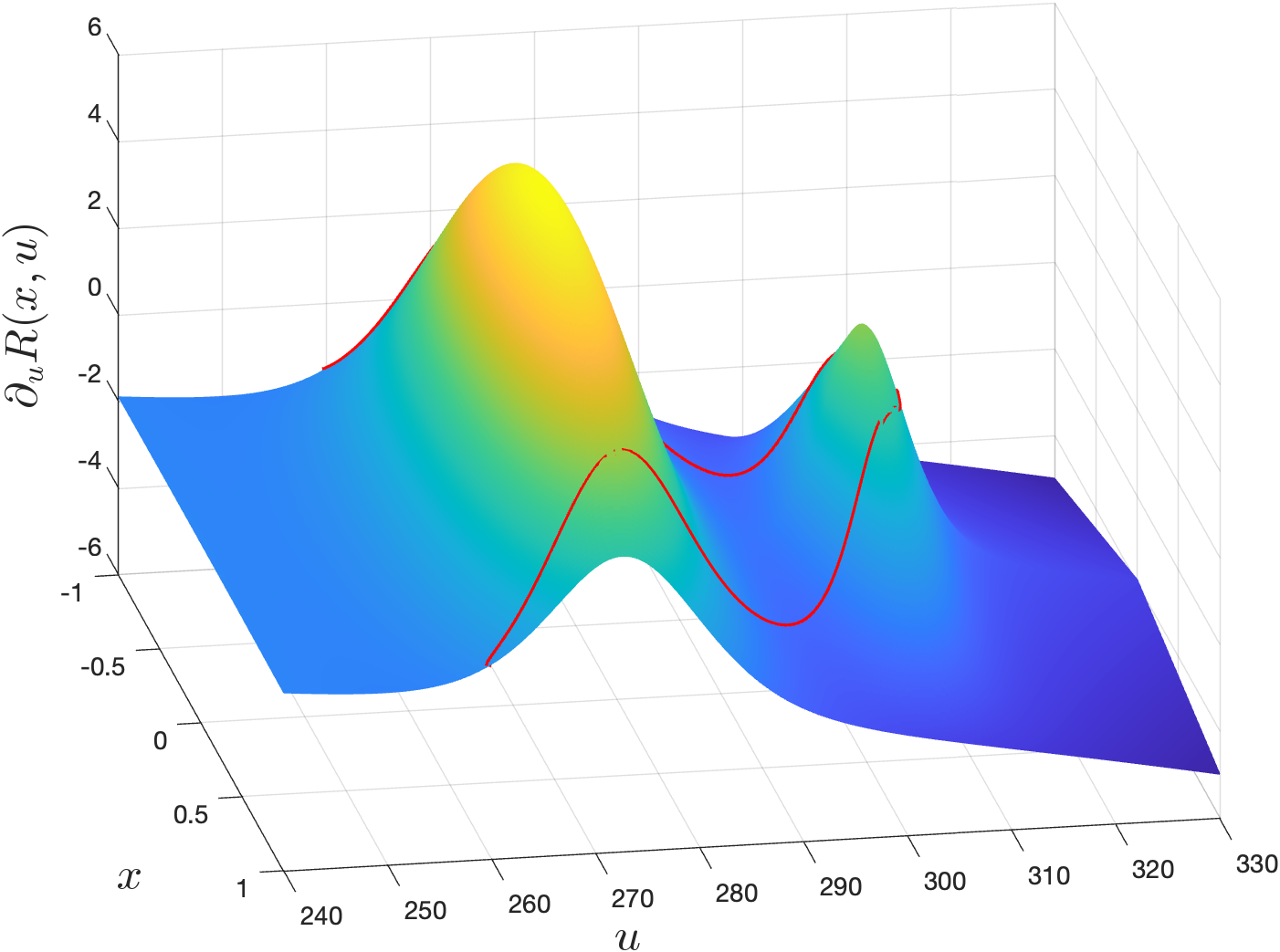}
\caption{$q = 11.4$}
\end{subfigure}%
}
\caption{Plot of the map \( (x,u) \mapsto \partial_u R(x,u) \) and the curve \( x \mapsto (x,u_W(x), \partial_u R(x, u_W^{(q)}(x)) \) in red, for different values of \( q \).}
\label{fig: 3D indicator sequence}
\end{figure}

\begin{figure}[!htb]
\centering
\begin{subfigure}{.5\textwidth}
  \centering
  \includegraphics[width=.95\linewidth]{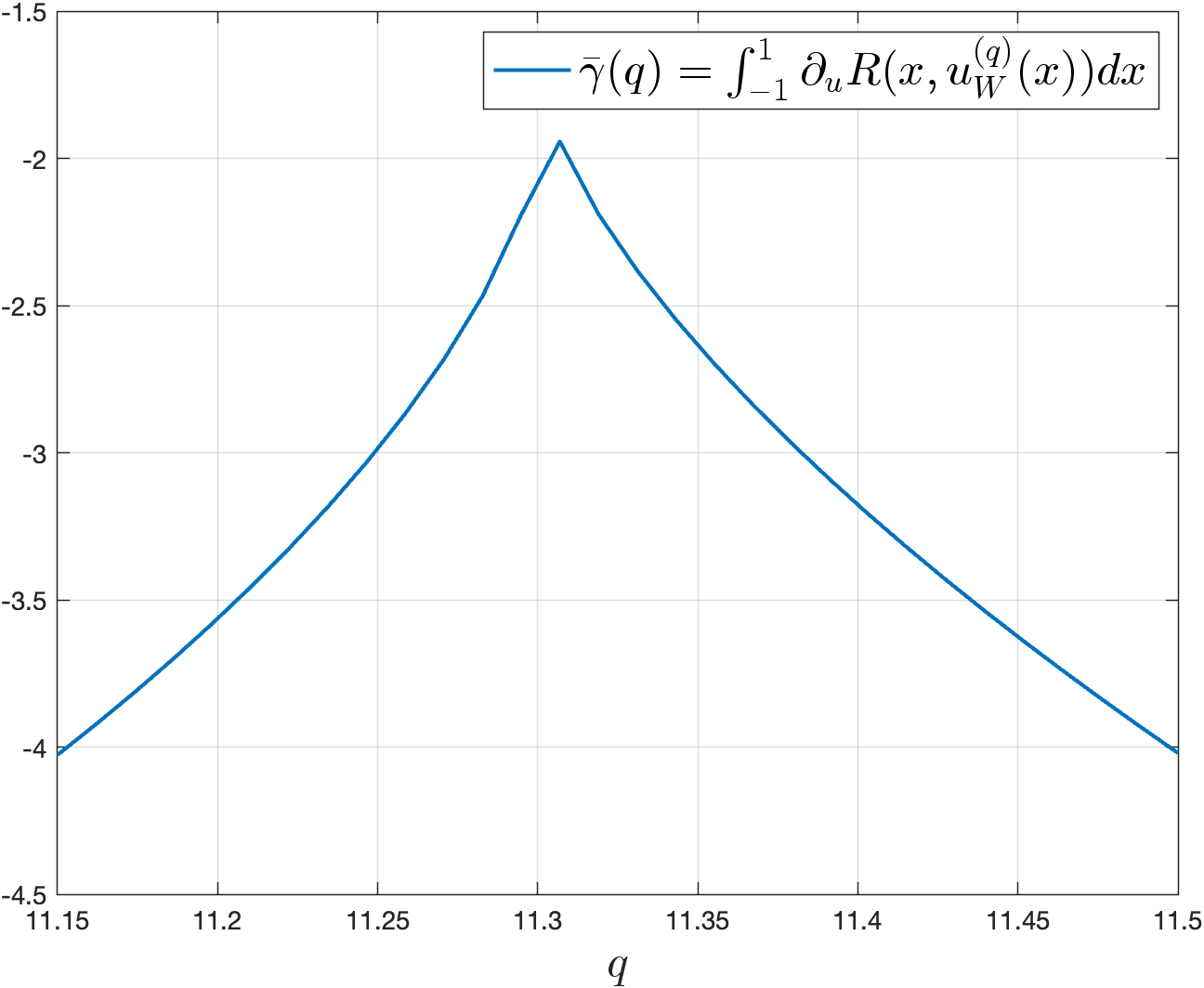}
  \caption{}
  \label{subfig: gamma medio}
\end{subfigure}%
\begin{subfigure}{.5\textwidth}
  \centering
  \includegraphics[width=.95\linewidth]{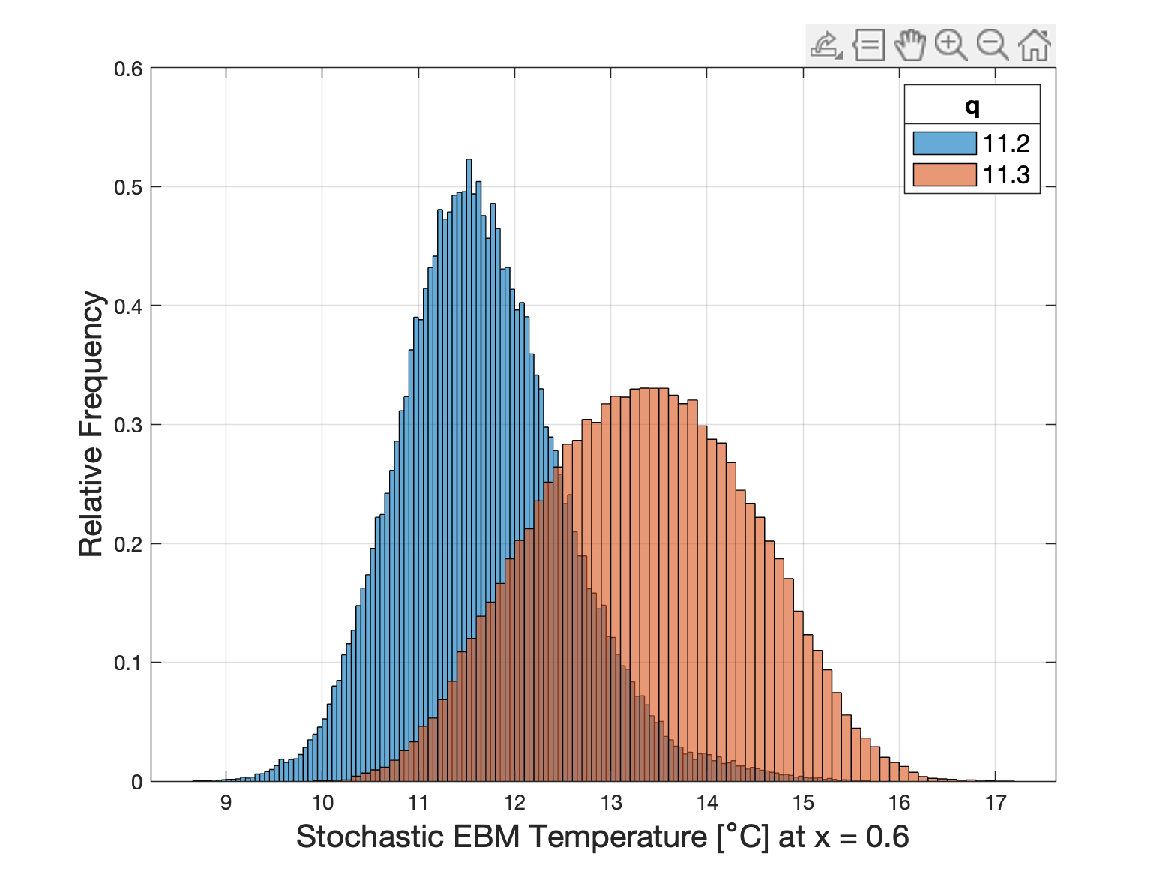}
  \caption{}
  \label{fig: histogram stochastic 1D EBM}
\end{subfigure}
\caption{(a) Average local stability indicator \( \Bar{\gamma}(q) = \int_{-1}^1 \partial_u R (x,u_W^{(q)}(x)) dx \). It increases, peaks around \( q \approx 11.3 \), and then decreases. (b) Histogram of the distribution of the solution of \( t \mapsto u(\Bar{x},t) \) for the stochastic EBM \eqref{eq: 1D EBM time variance} for \( q = 11.2 \) (blue) and \( q = 11.3 \) (red), at \( \Bar{x} = 0.6 \).}
\end{figure}

\begin{enumerate}
    \item[(i)] Figure \ref{fig: 3D indicator sequence} shows how the steady-state solution \( u_W^{(q)} \) approaches unstable areas, defined by points \( (x,u) \) where \( \partial_u R(x,u) > 0 \). For \( q_1 = 11.21 \), \( q_2 = 11.28 \), and \( q_3 = 11.4 \), the map \( (x,u) \mapsto \partial_u R(x,u_W(x)) \) is presented, with the curve \( x \mapsto (x,u_W(x), \partial_u R(x,u_W(x)) \) depicted in red. When the time variance peaks (i.e., \( q = q_2 \)), the steady-state solution values around the equator are close to the local maximum of \( \partial_u R \), approximately close to \( (x_M,u_M) = (0,305) \). For \( q = q_1 \) and \( q = q_3 \), the equatorial steady-state value is smaller and larger than \( u_M \), respectively.
    \item[(ii)] We consider the local mean stability indicator
\[
\Bar{\gamma}(q) = \int_{-1}^1 \gamma(x) dx =  \int_{-1}^1 \partial_u R(x, u_W^{(q)}(x)) dx.
\]
Its plot is shown in Figure \ref{subfig: gamma medio}. Although providing only average information, it exhibits behaviour qualitatively similar to that of \( \sigma^2_t \) for each fixed value of \( x \), peaking around \( q \approx 11.3 \).
\end{enumerate}

Finally, we show the distribution of the solution \( u(x,t) \) of the stochastic 1D EBM \eqref{eq: stochastic EBM} for the fixed space point \( x = 0.6 \) (similar results are observed for different space points). Figure \ref{fig: histogram stochastic 1D EBM} shows the distribution as a histogram, comparing two distinct values of \( q \). Blue depicts the results for \( q = 11.2 \), while red shows \( q = 11.3 \). From the plot, we deduce that our model captures the shift in the mean value and the increase in variance, correlating with the IPCC schematic climate change reproduction in Figure \ref{fig: IPCC} and weather observations in Modena in Figure \ref{fig: data Modena}.

\section{Conclusions}
\label{sec: conclusions}
In the first part of this work, we have presented a non-autonomous framework to describe the scale separation between weather, macroweather, and climate. According to Hasselmann's proposal, weather is conceptualised as arising from a set of deterministic equations, describing variables such as temperature on a fast timescale, typically from an hour to one day. Macroweather, on the other hand, encompasses variations that are neither as short-term as weather nor as long-term as climate. We assume that temperature on a macroweather timescale satisfies a stochastic equation, reflecting the most original aspect of Hasselmann's work. Additionally, at the macro weather timescale, we include a non-autonomous term to model the atmospheric $\text{CO}_2$ concentration, which evolves on a timescale of years. Finally, we identify climate with the invariant measure arising from the stochastic equation at the macro weather level.

In the second part, we have introduced a new 1D EBM on a macroweather timescale, which can predict the increase in the number of extreme weather events associated with climate change. The novelty of our model relies on the parametrisation of the non-linear space-dependent radiation budget. Indeed, we have inserted the presence of the SGE, a phenomenon typical of tropical areas that may lead to an instability in the OLR. Our model includes a non-autonomous term $q= q(t)$, that in light of the first part, we have considered constant, that is the effect of the $\text{CO}_2$ concentration on the radiation balance. We have recalled the basic mathematical properties of our model, such as the existence of steady-state solutions, using also numerical simulations.  Also, we have shown how the GMT of the steady state solution increases with increasing $\text{CO}_2$ concentration. Third, we have the stochastic version of our 1D EBM, obtained by perturbing our model with an additive space-time white noise. Being in the class of stochastic reaction-diffusion SPDE, it is possible to write explicitly the invariant measure as a Gibbs one. However, the presence of the non-linearity arising from the radiation budget prevents us from deducing more information from the invariant measure.

The most important results of our work are presented in Section \ref{sec: variance and extreme weather events increase }. There, we have exploited numerical simulations to study the changes in the invariant measure as the $\text{CO}_2$ concentration increases. To obtain this, we have performed numerical integration of the stochastic 1D EBM on a time interval of $500$ years and with the initial condition the warm steady-state solution $u_W$ of the deterministic 1D EBM, changing only $q$ in different runs of the simulation. Given $q$, for each space point $x \in [-1,1]$ we associate the extreme weather event frequency with the time variance $\sigma^2_t(x) $ of a trajectory of the stochastic 1D EBM at point $x$. We explain the spatial behaviour of $\sigma^2_t(x)$, which presents two local maximum points for $x \approx \pm 0.8$ and one for $x = 0$, combining heuristic reasoning with empirical indicators. The informal explanation is that the presence of the diffusion term forces the stable warm stationary solution, and the oscillations around it due to noise, to take on values that are not stable if we were to consider the ODE obtained by removing the diffusion term. These regions of temperature, which are locally unstable, correspond to the areas where the variance $\sigma^2_t$ is highest. We also motivated the behaviour of $\sigma^2_t (x) = \sigma^2_t{(q)}(x)$ with respect to $q$, at fixed $x$. The variance observed in this way increases to a maximum value and then decreases. This is explained by the increase in GMT due to $q$ and the presence of the diffusion term, leading the solution of the stochastic EBM to be increasingly in, and then out of, the region of instability due to the SGE. 

Lastly, we are aware that our model, although based on physical laws, is largely phenomenological. In some parts of our model, we have included simplifications of convenience, such as in the perturbation of the diffusion function, which makes the problem non-singular and thus easier to deal with mathematically; in other parts, such as in the parametrisation of the space-dependent and tropic-restored OLR, we have followed the principle of simplicity. Other choices would certainly have been possible, but the elementary nature of the model leads us to avoid choices that are too fine or complex.

Concerning open questions, an important one would be understanding the correct form of the noise term. We have assumed it to be additive and without correlation in space and time. This is an extreme simplification, which allows us to deduce certain properties of the invariant measure. But already Hasselmann in his seminal work proposed a different, i.e., multiplicative, type of noise. In addition to this, understanding how the properties of the model change if more physical processes are considered, such as advection, is a problem we would like to address in the future, as well as the extension of our work to a two-dimensional setting.

\appendix

\section{Appendix}
\label{sec: appendix}

\subsection{Numerical scheme}
\label{appendix: numerical scheme}

In this section, we describe the numerical method adopted to approximate the solutions of the stochastic PDE \eqref{eq: 1D EBM time variance}. We used an implicit Euler-Maruyama method, which is a small modification of the semi-implicit Euler-Maruyama method presented in \cite[Section 10.5]{Lord2014}.

First, it is worth recalling that the numerical experiments presented in Section \ref{sec: variance and extreme weather events increase } are all performed for a fixed value of $q$, and using $u_W = u_W^{(q)}$ as initial condition of the parabolic stochastic problem \eqref{eq: 1D EBM time variance}. The steady-state solution $u_W$ solves the elliptic problem \eqref{eq: elliptic PDE}. To numerically approximate it, we have applied the same finite difference scheme described in \cite[Appendix A]{DelSarto2024}.

Second, we move to describe the numerical method for the parabolic problem. We denote by
$
R(x,u;q) = R_a(x,u) -R_e(x,u;q)
$
the non-linear radiation budget, and the underline notation to denote a vector (e.g. $\underline{y} \in \mathbb{R}^4).$ We consider two uniform meshes for the spatial domain $[-1,1]$ and the time domain $[0,T]$, i.e. 
$$
 x_i = -1 + i \Delta x, \quad i=0,..., n, \quad \Delta x = \frac{2}{n},
$$
and
$$
 t_j =  j \Delta t, \quad j=0,..., m, \quad \Delta t= \frac{T}{m} = \frac{500}{m}.
$$
The number of points in the space and time mesh, respectively $n+1$ and $m+1$, are chosen in a way that $\Delta x = \Delta x = 0.01$.
Then, the solution to the problem can be approximated by considering the system 
\begin{equation*}
\begin{split}
C_T &\frac{u_{i,j+1}-u_{i,j}}{\Delta t} \\
&= 
\frac{u_{i-1,j+1}\kappa_{i-\frac{1}{2}}- u_{i,j+1}(\kappa_{i-\frac{1}{2}}+\kappa_{i+\frac{1}{2}})+u_{i+1,j+1}\kappa_{i+\frac{1}{2}}}{\Delta x^2} \\
&+ R(x_i,u_{i,j+1}) + \sqrt{\frac{\Delta t}{\Delta x}}   \sigma  z_{i,j}  \quad  0 \leq i \leq n, \, 0 \leq j \leq m-1,\\
0  &= \frac{u_{n+1,j} - u_{n-1,j}}{2 \Delta x} = \frac{u_{1,j} - u_{-1,j}}{2 \Delta x} \quad j = 1,...,m, \\
 u_{i,0} &= u_W(x_i), \quad i = 0,...,n,
\end{split}
\end{equation*}
where $u_{-1,j},u_{n+1,j}$ are ghost points, $\left(z_{i,j}\right)_{i,j}$ is a collection of independent identically distributed (i.i.d.) normal random variables,  and
$
u_{i,j}= u(x_i,t_j), \; \kappa_{i \pm \frac{1}{2}}= \kappa(x_{i \pm \frac{1}{2}}), \; x_{i \pm \frac{1}{2}}= x_i \pm \Delta x/2 .
$
Since $u_{1,j} = u_{-1,j}$ and $u_{n+1,j} = u_{n-1,j}$, the previous system of equation can be rewritten as
$$
\left(I_{n+1} - r A \right) \underline{u}^{j+1}  = I_{n+1} \underline{u}^j + \frac{\Delta T}{C_T} \underline{R}(\underline{u}^{j+1}) + \sqrt{\frac{\Delta t}{\Delta x}} \sigma \underline{Z}^j,
$$
where $I_{n+1}$ denotes the $(n+1) \times (n+1)$ identity matrix , $\underline{u}^j = \left(u_{0,j}, \cdots ,u_{n,j} \right)^T$, $r = \frac{\Delta t}{C_T \Delta x^2}$, $\underline{R}(\underline{y}) = \left(R(x_0, y_1), \cdots , R(x_n, y_{n+1} ) \right)^T$, $ \left( \underline{Z}^j \right)_j$ denotes a set of i.i.d. $\mathcal{N}(\underline{0}, I_{n+1})$ normal random vectors, and $A \in \mathbb{R}^{(n+1) \times (n+1)}$ is the tridiagonal matrix with diagonal $\underline{d} \in \mathbb{R}^{n+1}$, superdiagonal $\underline{d}_1 \in \mathbb{R}^n$, and subdiagonal $\underline{d}_{-1} \in \mathbb{R}^n$ given by
\begin{equation*}
    \begin{split}
        \underline{d}(i) &=\begin{cases}
            - ( \kappa_{-1/2}+ \kappa_{1/2}), & i = 1, \\
            - ( \kappa_{i-3/2} + \kappa_{i-1/2}),  &i = 2,...,n, \\
            - ( \kappa_{n-1/2}+ \kappa_{n+1/2}), & i = n+1,
        \end{cases}
        \\
        \underline{d}_1(i) &=\begin{cases}
            \phantom{-(} \kappa_{-1/2}+ \kappa_{1/2} \phantom{)}, \phantom{aa}  & i = 1, \\
             \quad \; \kappa_{i-1/2},  & i = 2,..., n, 
        \end{cases}
        \\
        \underline{d}_{-1}(i) &=\begin{cases}
             \phantom{-(} \kappa_{i-1/2}+ \kappa_{1/2} \phantom{)}, & i = 1,..., n, \\
             -( \kappa_{n-1/2}+\kappa_{n+1/2}), & i = n. 
        \end{cases}
    \end{split}
\end{equation*}

Thus, given the numerical approximation $\underline{u}^j$ at time $t_j$, to advance the scheme at time $t_{j+1}$ it is needed to solve the previous non-linear system of algebraic equations. To do this, we apply the Newton-Raphson Method (NRM), which we recall in the following. We consider the map $\underline{F}^j \colon \mathbb{R}^{n+1} \to \mathbb{R}^{n+1}$ defined as
$$
\underline{F}^j (\underline{y}) = \left( I_{n+1}  - r A \right) \underline{y} - \underline{u}^j  -\frac{\Delta T}{C_T} \underline{R}(\underline{y}) -\sqrt{\frac{\Delta t}{\Delta x}} \sigma \underline{Z}^j.
$$
To approximate the vector $\underline{y} \in \mathbb{R}^{n+1}$ such that $\underline{F}^j(\underline{y}) = 0$, the NRM consider the following sequence
\begin{equation*}
\begin{cases}
    \underline{y}^{k+1} &= \underline{y}^k - J_{F^j} (\underline{y}^k)^{-1} F^j(\underline{y}^k), \\
    \underline{y}^0 &= \underline{u}^j ,
\end{cases}
\end{equation*}
where $J_{\underline{F}^j} \in \mathbb{R}^{(n+1) \times (n+1)}$ is the Jacobian matrix of $\underline{F}^j$ with respect to $\underline{y}.$  Since the noise intensity $\sigma = 0.2$ is small, we expect that the choice  $\underline{y}^0 = \underline{u}_j$ is a good initial guess of the solution. The iteration of the NRM are stopped when $\norm{\underline{F}^j(\underline{y}^k} \leq 10^{-10}$.

\subsection{Spectral properties of the operator $\Tilde{A}$}
\label{appendix: sturm liouville operators}

To apply the invariant measure theory, the operator $\Tilde{A} = \lambda Id -A $ defined in Section \ref{sec: stochastic properties of the model} should satisfy the following assumptions (see \cite[Section 11]{DaPrato2006}):
\begin{enumerate}
    \item[(i)] $\Tilde{A}$ is self-adjoint and there exists $\omega >0$ such that
    $$
    \langle \Tilde{A}x,x \rangle \leq - \omega \abs{x}^2, \quad x \in D(A).
    $$
    \item[(ii)] There exists $\beta \in (0,1)$ such that $Tr \left[(-\Tilde{A})^{\beta -1} \right] <+\infty$.
\end{enumerate}
We denote by $(\lambda_n)_n$ the eigenvalues of $-A$, where 
\begin{equation}
\begin{split}
    D(A) &= \left \lbrace u \in H^2(-1,1) \; \mid \; u'(-1) = u'(1) = 0 \right \rbrace \\
    A u &=\left(\kappa(x) u' \right)'.
\end{split}
\end{equation}
Since $A$ is a Sturm-Liouville regular operator, classical results assure the existence of the real eigenvalues $\lambda_n$, and furthermore, it can be proved
$$
\lambda_1 < \lambda_2 < \cdots \cdots < \lambda_n < \cdots \to +\infty.
$$

First, to check hypothesis (i) it is thus sufficient to prove that 
$$
\lambda_n \geq 0, \quad \forall n.
$$
Indeed, by definition of eigenvalue, there exists an eigenfunction $v_n \in D(A)$ such that
$$
A v_n = - \lambda_n v_n.
$$
Multiplying the previous identity by $v_n$ and integrating over the domain $[-1,1]$, we get
$$
\int_{-1}^1 \left[\kappa(x) v'_n(x) \right]' v_n(x) dx = - \lambda_n \int_{-1}^1 v_n(x)^2 dx.
$$
Performing an integration by parts on the left-hand side, we deduce
$$
- \int_{-1}^1 \kappa(x) \left( v_n'(x) \right)^2 dx = - \lambda_n \int_{-1}^1 v_n(x)^2 dx.
$$
In conclusion,
$$
\lambda_n = \frac{ \int_{-1}^1 \kappa(x) \left( v_n'(x) \right)^2 dx }{\int_{-1}^1 v_n(x)^2 dx},
$$
and our claim follows from the fact that $\kappa(x) >0 $ on $[-1,1]$.

Second, the hypothesis (ii) is a consequence of the following asymptotic estimate for regular Sturm-Liouville problems (\cite[Section 4]{Fulton1994}): there exist $B>0$ and $n_0 >0$ such that 
$$
\left[\frac{\left( n- \frac{1}{2} \right) \pi }{B} \right]^2 < \lambda_n < \left[ \frac{\left( n + \frac{1}{2}\right) \pi}{B} \right]^2, \quad \forall n \geq n_0.
$$

\section*{Code and data availability}
This work does not include any externally supplied code. All material in the text and figures was produced by the authors using standard mathematical and numerical analysis tools. The code for the numerically simulation of the 1D EBM is available at Zenodo (\href{https://doi.org/10.5281/zenodo.11609953}{https://doi.org/10.5281/zenodo.11609953}). The Modena temperature data were provided by the Geophysical Observatory of Modena, University of Modena and Reggio Emilia, Italy (\href{www.ossgeo.unimore.it}{www.ossgeo.unimore.it}, \cite{Lombroso2008}) and are available upon request from the corresponding author.

\section*{Acknowledgments}
We acknowledge fruitful discussions with Paolo Bernuzzi and Giulia Carigi. The research of G.D.S. is supported by the Italian national interuniversity PhD course in sustainable development and climate change. The research of F.F. is funded by the European Union (ERC, NoisyFluid, No. 101053472). Views and opinions expressed are however those of the authors only and do not necessarily reflect those of the European Union or the European Research Council. Neither the European Union nor the granting authority can be held responsible for them.

%\nocite{*}
\bibliographystyle{alpha}
\bibliography{aipsamp}

\end{document}